\documentclass[
reprint,
superscriptaddress,
floats,
twocolumn,
 amsmath,amssymb,aps,
 showpacs,
prd
]{revtex4-2}

\usepackage{graphicx}
\usepackage{dcolumn}
\usepackage{bm}
\usepackage{subfigure}
\usepackage{enumerate}
\usepackage{multirow}
\usepackage{makecell}
\usepackage{color}
\usepackage{tabularx}
\usepackage{amsmath}
\usepackage{amssymb}
\usepackage{datetime}
\usepackage{txfonts}
\usepackage{iitem}
\usepackage{array}
\usepackage{siunitx}
\usepackage{url}
\usepackage{footnote}
\usepackage{booktabs}

\newcommand{\tabincell}[2]{\begin{tabular}{@{}#1@{}}#2\end{tabular}}
\newcommand{\PD}{$p\to \bar{\nu} K^+$\ }
\newcommand{\AN}{atmospheric $\nu$\ }
\newcommand{\KPLUS}{$K^+$}

\begin{document}


\title{JUNO Sensitivity on Proton Decay $p \to \bar{\nu} K^+$ Searches}
\author {Angel Abusleme}
\affiliation {Pontificia Universidad Cat\'{o}lica de Chile, Santiago, Chile}

\author {Thomas Adam}
\affiliation {IPHC, Universit\'{e} de Strasbourg, CNRS/IN2P3, F-67037 Strasbourg, France}

\author {Shakeel Ahmad}
\affiliation {Pakistan Institute of Nuclear Science and Technology, Islamabad, Pakistan}

\author {Rizwan Ahmed}
\affiliation {Pakistan Institute of Nuclear Science and Technology, Islamabad, Pakistan}

\author {Sebastiano Aiello}
\affiliation {INFN Catania and Dipartimento di Fisica e Astronomia dell Universit\`{a} di Catania, Catania, Italy}

\author {Muhammad Akram}
\affiliation {Pakistan Institute of Nuclear Science and Technology, Islamabad, Pakistan}

\author {Fengpeng An}
\affiliation {East China University of Science and Technology, Shanghai, China}

\author {Qi An}
\affiliation {University of Science and Technology of China, Hefei, China}

\author {Giuseppe Andronico}
\affiliation {INFN Catania and Dipartimento di Fisica e Astronomia dell Universit\`{a} di Catania, Catania, Italy}

\author {Nikolay Anfimov}
\affiliation {Joint Institute for Nuclear Research, Dubna, Russia}

% ...

\collaboration{JUNO Collaboration}
\date{\today}

\begin{abstract}

The Jiangmen Underground Neutrino Observatory (JUNO) is a large liquid scintillator detector designed to explore many topics in fundamental physics.
In this paper, the potential on searching for proton decay in \PD mode with JUNO is investigated. 
The kaon and its decay particles feature a clear three-fold
coincidence signature that results in a high efficiency for identification. Moreover, the excellent energy resolution of JUNO permits to suppress the sizable background caused by other delayed signals. Based on these advantages, the detection efficiency for the proton decay via \PD is $36.9\%\pm4.9\%$ with a background level of $0.2\pm 0.05({\rm syst})\pm 0.2({\rm stat})$ events after 10 years of data taking. The estimated sensitivity based on 200 kton-years exposure is $9.6 \times 10^{33}$ years, which is competitive with the current best limits on the proton lifetime in this channel and complementary using different detection technologies.

\end{abstract}

\maketitle

\section{Introduction} \label{Sec.1}
To explain the observed cosmological matter-antimatter asymmetry, the baryon number $B$ violation is one of three basic ingredients for an initially symmetrical Universe \cite{Sakharov:1967dj}. 
The baryon number is necessarily violated in the Grand Unified Theories (GUTs) \cite{Georgi:1974sy,Nath:2006ut}, which can unify the strong, weak and electromagnetic interactions into a single underlying force at a scale of $M_\mathrm{GUT}\simeq 2 \times 10^{16}$~GeV. 
A general prediction of the GUTs is proton decay. 
However, no experimental evidence of proton decay, $B$-violating neutron decay and neutron-antineutron oscillation has been found \cite{PDG}. 
Fortunately, the new generation of underground experiments JUNO \cite{JUNO,JUNO_PPNP}, Hyper-Kamiokande \cite{Hyper-K} and DUNE \cite{DUNE} with huge target masses and different detection technologies will continue to search for proton decay and test the GUTs.

Among many possible proton decay modes \cite{PDG}, $p\to e^+ \pi^0$ and \PD are the two dominant ones predicted by a majority of GUTs.
The first one is expected to be the leading mode in many GUTs, particularly in those non-supersymmetric GUTs which typically predict the lifetime of proton to be about $10^{35}$ years \cite{Babu}. 
In comparison, the decay mode \PD is favored by a number of supersymmetric GUTs. 
For these two decay modes, best measured upper limits of proton partial lifetime are $\tau/B(p \to e^+ \pi^0) > 2.4 \times 10^{34}$ years \cite{Takenaka:2020vqy} and $\tau/B(p \to \bar{\nu} K^+) > 5.9 \times 10^{33}$ years \cite{Abe:2014mwa} at $90\%$ C.L. from the Super-Kamiokande (Super-K) experiment, which is a water Cherenkov detector. 

Compared to the water Cherenkov detectors, a liquid scintillator (LS) detector has a distinct advantage in detecting the proton decay mode \PD \cite{Svoboda,LENA, KamLAND,JUNO}. 
The present paper plans to investigate the sensitivity of the future LS detector, JUNO. Here, the decay will give rise to a three-fold coincidence feature in time, which is usually composed of a prompt signal by the energy deposit of $K^+$, a short-delayed signal ($\tau=12.38$ ns) by the energy deposit of decay daughters of $K^+$ and a long-delayed signal ($\tau=2.2$ $\mu$s) by the energy deposit of the final Michel electron. 
Using the time-correlated triple coincidence, the JUNO detector can effectively identify the \PD and reject the atmospheric neutrino backgrounds \cite{KamLAND}.

Preliminary studies have given a rough estimation of the sensitivity of JUNO to the proton decay mode \PD  \cite{JUNO}. 
In this paper, the JUNO potential based on a detailed detector performance has been studied in Monte Carlo (MC) simulation.
Sec.~\ref{Sec.2}  briefly introduces the JUNO detector and its expected performance. 
In Sec.~\ref{Sec.3}, the MC simulation of \PD and the \AN backgrounds will be described. 
In Sec.~\ref{Sec.4}, the multi-pulse fitting method and other selection criteria to discriminate \PD from the backgrounds are investigated. 
We will present the expected sensitivity of JUNO to the \PD in Sec.~\ref{Sec.5}. 
Finally, a conclusion is given in Sec.~\ref{Sec.6}.

\section{JUNO Detector} \label{Sec.2}

JUNO is a multi-purpose neutrino observatory under construction in South China.
As a low background observatory, it has a vertical overburden of 700 m rock (1800~m.w.e) to shield the detector from cosmic muons.
Its central detector (CD) is a 12 cm thick acrylic sphere with a diameter of 35.4 m, filled with 20 kton LS.
The CD is immersed in a cylindrical water pool and supported by a stainless steel lattice structure.
Besides, the CD is instrumented with 17612 20-inch PMT (LPMT) and 25600 3-inch PMT (SPMT) which are uniformly distributed outside the acrylic sphere. 
5000 of the LPMT are dynode (DYN) PMT produced by Hamamatsu Photonics K.K., while the remaining LPMT are Micro Channel Plate (MCP) PMT manufactured by North Night Vision Technology Co. Ltd. (NNVT) \cite{MCPPMT:2020}. 
Their transit time spread (TTS) are 1.1 ns and 5.0 ns in $\sigma$, respectively, according to the result of the PMT mass test \cite{Wen:2019sik}.
The total photocathode coverage of the LPMT will be around $75\%$. 
The SPMT, which contribute another $2.5\%$ photocathode coverage, are also deployed to serve as an additional independent calorimeter. 
The TTS ($\sigma$) of SPMT has been measured to be around 1.5 ns \cite{Cao:2021wrq}.
For each MeV energy deposition in LS when detecting the low energy events, around $1.3\times 10^{3}$ photonelectrons (PE) are expected to be received by the LPMT. 

A VETO system, including Top Tracker (TT) detector and water Cherenkov PMT system, is designed to prevent the influence of cosmic muons. 
The TT detector is a plastic scintillator detector complex which partly covers the water pool and the CD, which helps reject the cosmic muons passing it. 
The water Cherenkov PMT system is assembled on the outer surface of the stainless steel lattice structure and measures the Cherenkov light produced by the cosmic muons passing the water pool. The rejection ratio of cosmic muons is estimated to be more than 99\%.

\section{Simulation} \label{Sec.3}

To understand the behavior of \PD and to discriminate them from the backgrounds in JUNO detector, a Monte Carlo simulation has been performed which is composed of two steps, the generator production and detector simulation.
The generator of \PD and its backgrounds is produced with GENIE (version 3.0.2) \cite{GENIE}, in which the primary processes of \PD and the \AN interactions in LS are simulated.
The detector simulation, which is the simulation of the final states of \PD and \AN interaction in the JUNO detector, is processed in SNiPER \cite{SNIPER} which is a Geant4 \cite{Agostinelli:2002hh} based simulation software developed by the JUNO collaboration. 
All the related optical processes, including the quenching effect, are considered.
The profiles of the LS, including the fluorescence times can be found in Ref. \cite{LStimeprofile}. 
In total, \SI{10}{k} \PD (PD) events and \SI{160}{k} \AN events are simulated with vertex positions uniformly distributed over the whole LS volume.

This study does not yet use a full event reconstruction of energy, position and hit time information. 
Instead, they are smeared according to the expectation from the detector Monte Carlo and used as the input to our further analysis. 
The visible energy ($E_\text{vis}$) is the energy deposition reconstructed from the number of PE received by the LPMT. 
For a conservative consideration, it is smeared by ${3\%}/{\sqrt{E_\text{vis} ({\rm MeV})}}$ when the energy deposition is smaller than 60 MeV, and a resolution of $1\%$ when greater \cite{CalibStrategy}.
The position of the event is described with the center of energy deposition position, which is the averaged position weighted by the energy deposition each time. 
It is smeared by a Gaussian distribution with resolution of 30~cm. 
In this study, the detected times of the photons hit on the cathode of the SPMT are collected to form a hit time spectra for each event, after the correction of photon time-of-flight (TOF) relative to the reconstructed deposition center. 
TTS of SPMT are set randomly according to the measurement results introduced in Sec. \ref{Sec.2}.
The reason for not using the LPMT will be introduced in Sec. \ref{sec:PD}.

\subsection{Proton Decay}\label{sec:PD}

\begin{figure}[h]
\centering
\includegraphics[width=8cm]{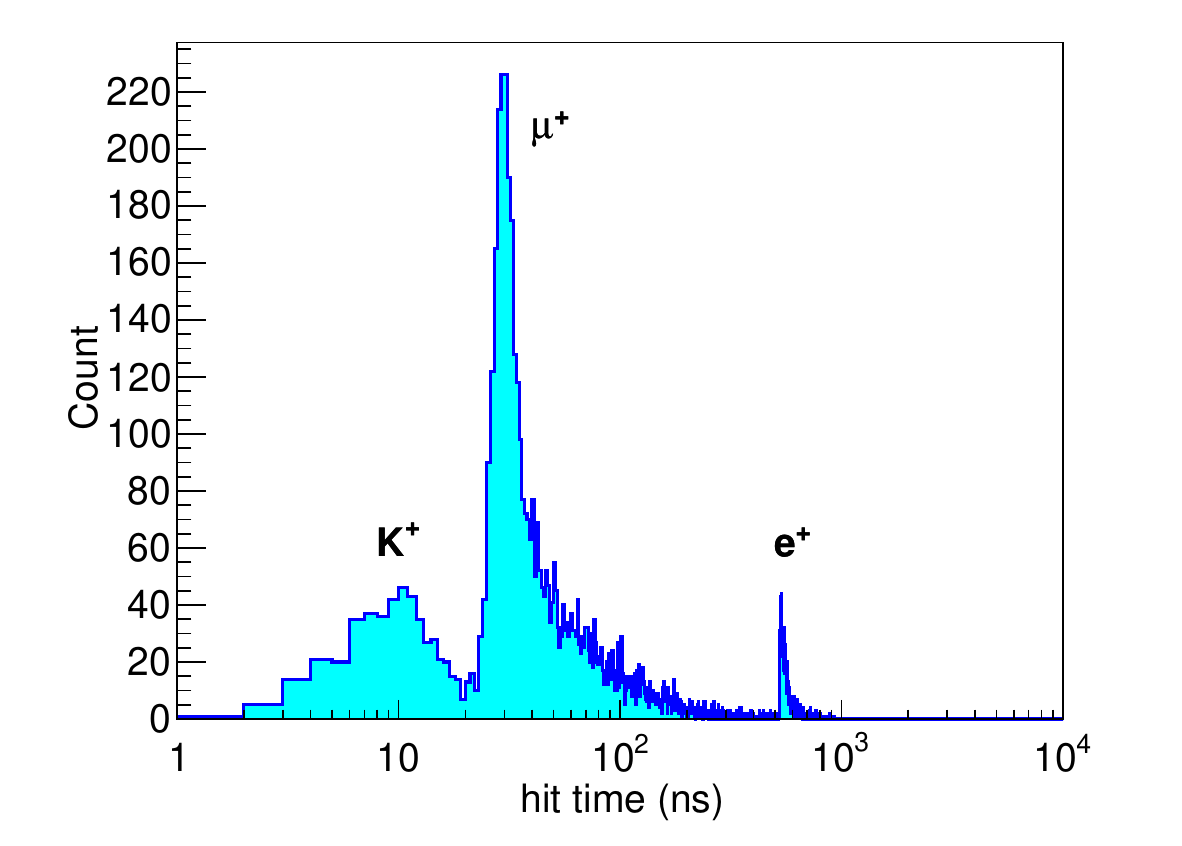}
\caption{\label{fig:TimeSpec} Illustration of the hit time spectrum of a typical \PD event, containing the signals of $K^+$, the decay daughter of $K^+$ ($\mu^{+}$ in this event) and the Michel electron.}
\end{figure}

Based on the JUNO LS components, the initial proton of \PD may come from free protons (in Hydrogen) or bound protons (in Carbon). 
In free proton decay, the final states $\bar{\nu}$ and $K^+$ have fixed kinetic energies of 339~MeV and 105~MeV, respectively.
According to a toy MC simulation with the corresponding monochromatic $K^+$ in the JUNO detector, it is found that $92.4\%$ of $K^+$ will deposit all of their kinetic energy within 1.2~ns, which means a signal can be found in the hit time spectrum immediately.
Then, these $K^+$ will stay at rest until decaying into their daughter particles after an average of 12.38 ns. 
The $K^+$ has six main decay channels.
The most dominant channels are $K^+ \to \mu^+ \nu_\mu$ and $K^+ \to \pi^+ \pi^0$ with branching ratios of $63.56 \%$ and $20.67 \%$, respectively \cite{PDG}.
In the first channel, the produced $\mu^+$ has a kinetic energy of 152 MeV and decays to a Michel electron with a lifetime of about 2.2 $\mu s$.
The produced $\pi^0$ and $\pi^+$ in the second channel will decay into two gammas, a $\mu^+$ plus a $\nu_\mu$, respectively, and consequently produce a Michel electron. 
All daughter particles will deposit their kinetic energies immediately and give a second signal.
After the TOF correction, the hit time spectrum of the $K^{+}$ and decay particles will form an overlapping double-pulse pattern.
Given the relatively long lifetime of the muon, a later third pulse from the Michel electron, as a delayed feature of $p\to \bar{\nu} K^+$, will be found on the hit time spectrum.
This triple coincidence as introduced in Sec.~\ref{Sec.1} is one of the most important features to distinguish a \PD event from the backgrounds.
This triple coincidence is illustrated in Fig.~\ref{fig:TimeSpec}.

\begin{figure}[ht]
    \subfigure[Waveform of a LPMT acquired from the SNiPER electronics simulation.]{
        \includegraphics[height=5cm]{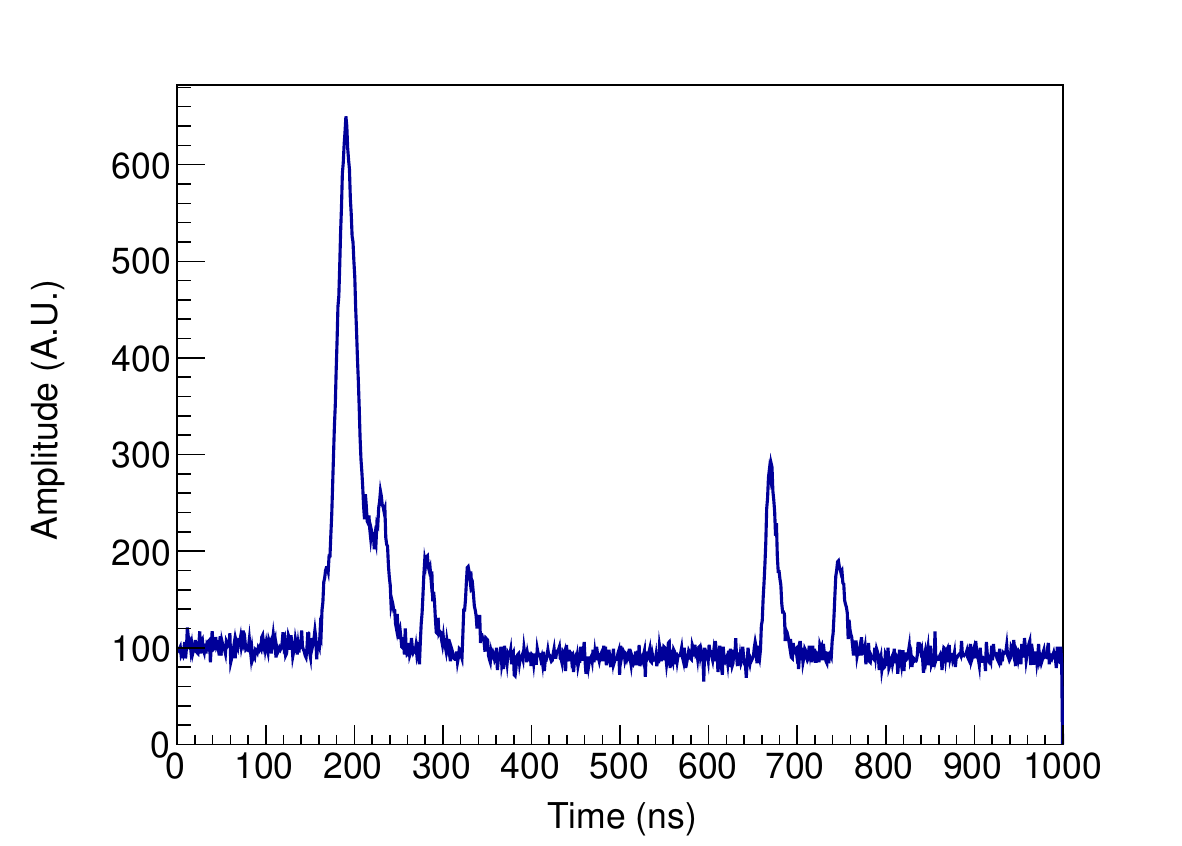}
        \label{fig:singlePMT}
    }
    \subfigure[Comparison of the LPMT waveform and SPMT hit time output from a typical \PD event after TOF correction.]{
        \includegraphics[height=5cm]{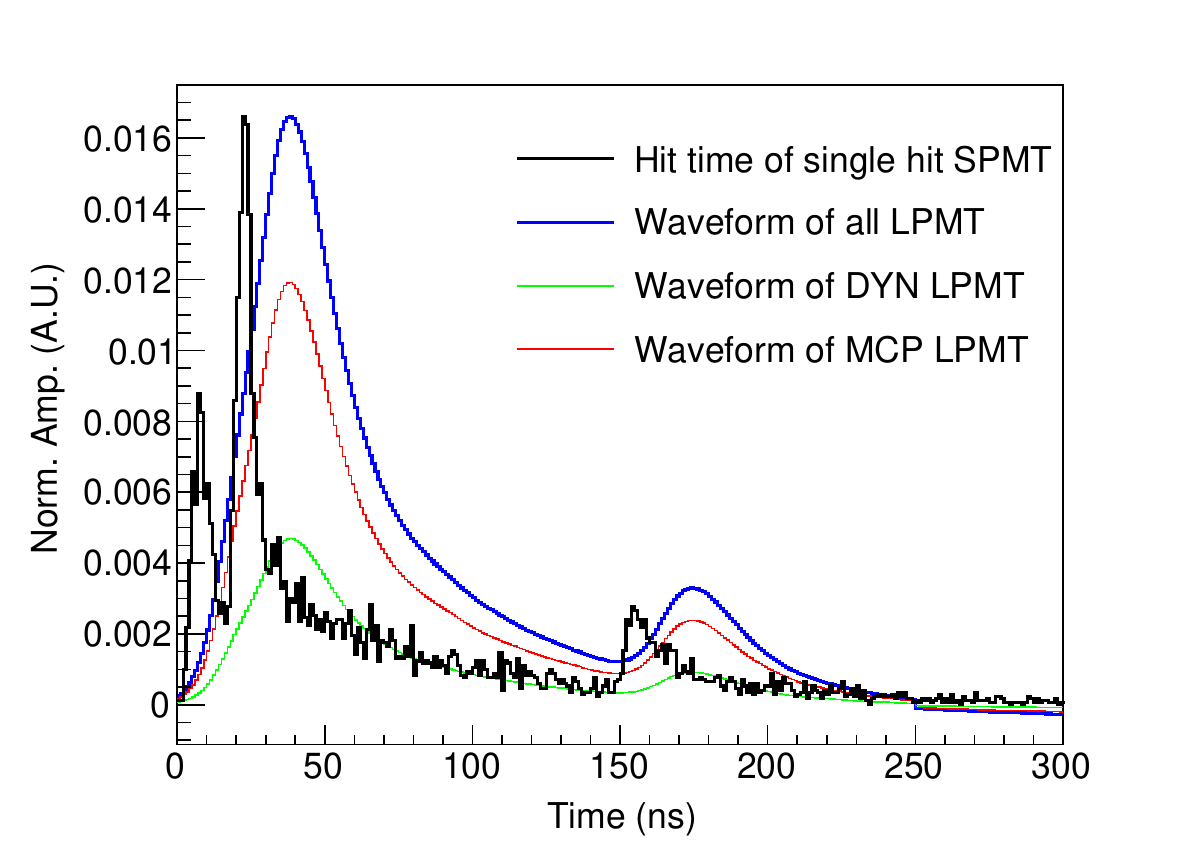}
        \label{fig:3pulses}
    }
\caption{ \label{fig:PMTPerformance} Simulated PMT output of a typical \PD event. The total visible energy of this event is 275 MeV and the \KPLUS decays at 13.7 ns after it's born. Photon hit time reconstruction is not easy to achieve when using LPMT to detect a hundreds-of-MeV event. So the SPMT is used for hit time spectrum collection. More details can be found in the text.}
\end{figure}

As introduced in Sec.~\ref{Sec.2}, both the LPMT and SPMT are used in JUNO. 
However, as shown in Fig.~\ref{fig:PMTPerformance}, they have different performances on the hit time spectrum collection. 
When a LPMT is triggered by a hit, the waveform will be digitized and recorded by the electronics. 
Then, the hit time reconstruction (from the waveform to the hit time of each PE) will be carried out to get the hit time spectrum.
For low energy events such as the inverse $\beta$ decay (IBD), the hit time reconstruction is possible since only a few photons could be received by most LPMT.
However, a typical \PD event usually has an energy deposition of more than 200 MeV. 
In this case, many PEs would be received by the LPMT in a few tens of ns (as shown in Fig.~\ref{fig:singlePMT}) and the hit time reconstruction would be difficult. 
As shown in Fig.~\ref{fig:3pulses}, the overlapping of the first two pulses of the triple coincidence time feature would be smeared if the hit time reconstruction is not carried out. 
Thus, the LPMT are not used to collect the hit time spectrum in this study.
In comparison, considering that the receiving area of SPMT is around 1/40 times that of LPMT, most SPMT will work in single hit mode in which the SPMT is usually hit by at most only one PE. 
Advantageously, the triple coincidence time feature of \PD could be preserved well. Thus, only the SPMT in single hit mode are used in this study to collect the hit time spectrum.

The protons bound in Carbon nuclei will be influenced by nuclear effects \cite{Abe:2014mwa}, including the nuclear binding energy, Fermi motion and nucleon-nucleon correlation. 
The kinetic energies of the produced $K^+$ are smeared around 105 MeV which is relative to that in the free proton case. 
In addition, the $K^+$ kinetic energy will also be changed by the final state interactions (FSI). 
Before the $K^+$ escapes from the residual nucleus, it may interact with the spectator nucleons and knock one of them out of the remaining nucleus. 
It can also exchange its charge with a neutron and turn into $K^0$ via $K^+ + n \to K^0 +p$. 
Furthermore, the de-excitation of the residual nucleus will produce $\gamma$s, neutrons or protons etc. 
Obviously, the FSI and de-excitation processes will change the reaction products, which are crucial to our later analysis. 
 
The GENIE generator (version 3.0.2) \cite{GENIE} is used to model these nuclear effects.
Some corrections have been made to the default GENIE. 
Firstly, the nuclear shell structure is taken into account which is not included in the default nuclear model of GENIE.
A spectral function model, which provides a 2-dimensional distribution of momentum $k$ and removal energy $E_R$ for protons in $^{12} {\rm C}$, is applied to describe the initial proton states \cite{Benhar:2005dj}. 
Then, the initial proton energy is determined by $E_p = m_p - E_R$ where $m_p$ is the mass of a free proton. 
In this case, about $2.2\%$ of the protons from $^{12} {\rm C}$ cannot decay into $\bar{\nu}$ and $K^+$ since the corresponding proton invariant mass is smaller than the $K^+$ mass \cite{Hu:2021xjz}. 

\begin{figure}[h]
\centering
\includegraphics[width=7cm]{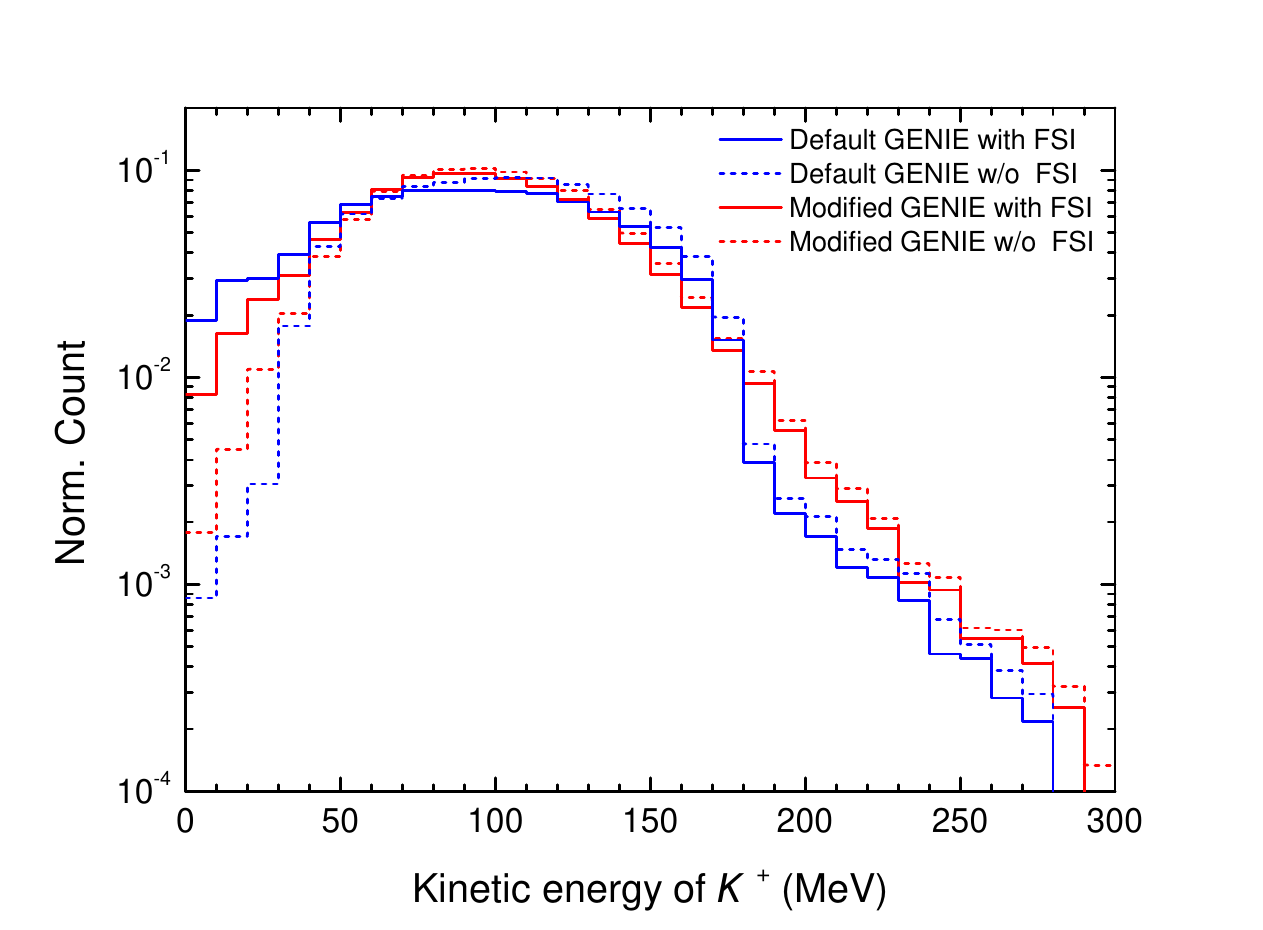}
\caption{\label{fig:KaonKinetic} The $K^+$ kinetic energy distributions for \PD in $^{12}$C with (solid line) and without (dashed line) the FSI from the default (blue) and modified (red) GENIE. }
\end{figure}

Secondly, we turn on the hadron-nucleon model in GENIE. The default GENIE uses the hadron-atom model to evaluate the FSI, which costs less time but does not include the $K^+ + n \to K^0 +p$ interaction.
Meanwhile, we modify the target nucleon energy and the binding energy with $m_p - E_R$ (or $m_n - E_R$) and $E_B = E_R - k^2/(2 M_{^{11}\rm{B}})$ \cite{Bodek:2018lmc}, respectively.
In addition, the fraction of $K^+$-nucleon charge exchange and elastic scattering interactions is corrected in terms of the numbers of spectator protons and neutrons in the remaining nucleus.
With all these modifications, we finally got a distribution of \KPLUS kinetic energies as shown in Fig.~\ref{fig:KaonKinetic}. The charge exchange probability is about $1.7\%$ for \PD in $^{12}$C according to the result of the modified GENIE. 

Thirdly, all the residual nuclei in the default GENIE are generated in the ground state, thus no de-excitation processes are taken into account.
The TALYS (version 1.95) software \cite{Koning:2012zqy} is then applied to estimate the de-excitation processes due to the excitation energy $E_x$.
The $E_x$ of the residual nucleus can be calculated through $E_x =M_{inv}-M_R$, where $M_{inv}$ and $M_R$ are the corresponding invariant mass and static mass, respectively.
For \PD in $^{12} {\rm C}$, $^{11} {\rm B^*}$,  $^{10} {\rm B^*}$ and $^{10} {\rm Be^*}$ account for $90.9 \%$, $5.1 \%$ and $3.1 \% $ of the residual nuclei, respectively. 
Among these residual nuclei, the $^{10} {\rm B^*}$ and $^{10} {\rm Be^*}$ come from the final state interactions between $K^+$ and one of the nucleons in $^{11} {\rm B^*}$.
The de-excitation modes and corresponding branching ratios of the residual nuclei $^{11} {\rm B^*}$, $^{10} {\rm B^*}$ and $^{10} {\rm Be^*}$ have been reported in Ref. \cite{Hu:2021xjz}. 

According to the result, many de-excitation processes could produce neutron.
In the case of a $s_{1/2}$ proton decay, the dominant de-excitation modes of $^{11} {\rm B^*}$ states, including $n + {^{10} {\rm B}}$, $n + p +  {^{9} {\rm Be}}$, $ n + d + {^{8} {\rm Be}}$, $n + \alpha +   {^{6} {\rm Li}}$, $2n +p + {^{8} {\rm Be}}$, will contribute to a branching ratio of $45.8 \%$ \cite{Hu:2021xjz}. 
About $56.5 \%$ of highly excited $^{11} {\rm B^*}$ states can directly emit one or more neutrons from their exclusive de-excitation modes. 
In addition, the non-exclusive de-excitation processes, and the de-excitation modes of $d +  {^{9} {\rm Be}}$ and $d + \alpha +  {^{5} {\rm He}}$, can also produce neutrons \cite{Hu:2021xjz}.
Most of these neutrons will give a 2.2 MeV $\gamma$ from the neutron capture in the JUNO LS, which will influence the setting of the criteria (introduced in Section. \ref{Sec:4.2}).

\subsection{Backgrounds} \label{sec:AN}

The dominant backgrounds of \PD are caused by \AN and cosmic muon since the deposited energy of \PD events are usually larger than 100~MeV. 
The cosmic muons come from the interaction of cosmic rays and the atmosphere. 
The produced cosmic muons usually have a very high energy and produce obvious Cherenkov light when passing through the water pool outside JUNO CD. With the VETO system, JUNO is expected to discriminate more than 99\% of the cosmic muons. 
The muons not detected by the VETO system usually clip the corner of the water pool with a very low energy deposited and few Cherenkov photons produced, and therefore escape from the watch of the VETO system. 
Thus, most VETO survived cosmic muons leave no signal in the CD and will not be background for \PD observation. 
For those muons that are VETO survived, entering and leaving signals in the CD, the energy deposition processes are mainly caused by the energetic primary muon. 
Consequently, with the visible energy, VETO and volume selection, as well as the expected triple coincidence feature selection, this type of background is considered to be negligible.
Therefore, the background mainly discussed in this paper is from \AN events. 

The expected number of observed \AN events is calculated with the help of the \AN fluxes at the JUNO site \cite{Honda:2015fha}, the neutrino cross sections from the GENIE \cite{GENIE} and the best-fit values of the oscillation parameters in the case of the normal hierarchy \cite{PDG}. The JUNO LS detector will observe 36k events in ten years. 
We use GENIE in its default configuration to generate 160 k \AN events, which corresponds to 44.5 years of JUNO data taking or 890 kton-years exposure mass.
Each \AN event has a weight value, which indicates the possibility of this event occurring for JUNO's 200 kton-years exposure considering the neutrino oscillation.
Then, these \AN events are simulated in SNiPER as our sample database. 

The \AN events can be classified into the following four categories \cite{Formaggio:2013kya}: 
the charged current quasi-elastic scattering (CCQE), the neutral current elastic scattering (NCES), the pion production and the kaon production. 
The categories and their ratios are shown in Table \ref{tab:ANcomponents}. 
The most dominant backgrounds in the energy range of \PD (Sub-GeV) are formed by elastic scattering, including the CCQE and the NCES events. 
The final states of the elastic scattering events usually deposit all their energy immediately and eventually followed by a delayed signal.
Consequently, requiring a triple coincidence feature effectively suppresses these two categories of backgrounds.

\begin{table*}[ht]
\centering
\caption{\label{tab:ANcomponents} The categories of \AN backgrounds. The data are summarized based on the result of GENIE and SNiPER.}
\begin{tabular}{c|c|c|c|c}
\Xhline{1.2pt}
Type & Ratio (\%) & \tabincell{c}{Ratio with $E_\text{vis}$ in \\ $[100~\rm{MeV},600~\rm{MeV}](\%)$} & Interaction & \tabincell{c}{Signal\\ characteristics} \\ \hline
NCES     & 20.2  & 15.8  & 
\tabincell{c}{$\nu+n \to \nu+n$\\$\nu+p \to \nu+p$}  & Single Pulse     \\ \hline
CCQE & 45.2  & 64.2  & \tabincell{c}{$\bar{\nu_{l}}+p \to n+l^{+} $\\$\nu_{l}+n\to p+l^{-} $} & Single Pulse     \\ \hline
Pion Production    & 33.5  & 19.8  & \tabincell{c}{$\nu_l +p \to l^{-}+p+\pi^{+}$\\$\nu+p\to \nu+n+\pi^{+}$} & \tabincell{c}{Approximate Single Pulse\\(Second pulse too low)}\\ \hline
Kaon Production    & 1.1   & 0.2   & \tabincell{c}{$\nu_l +n\to l^{-}+\Lambda+K^{+}$\\$\nu_l +p\to l^{-}+p+K^{+}$} & Double Pulse     \\ \Xhline{1.2pt}  
\end{tabular}
\end{table*}

Another significant background is CC and NC pion production, which is caused by single pion resonant interactions and coherent pion interactions, respectively. 
The produced pions will decay into muons with an average time of 26 ns. 
These muons, together with those produced in CC pion production, will consequently produce Michel electrons. 
It can be found that pion-production events would feature a triple coincidence in time similar to the search for \PD.
However, the muon contributed to the second pulse of the triple coincidence has kinetic energy of 4 MeV which is too small compared to the total energy deposition.

The \AN interactions with pion production have a larger possibility to produce the accompanying nucleons. 
Some of the created energetic neutrons have a small probability to propagate freely for more than 10 ns in the LS. 
In this case, the neutron interaction can cause a sufficiently large second pulse. 
Therefore, pion production events with an energetic neutron, e.g. $v + p \to v + n + \pi^+$, can mimic the signature of \PD. 
In fact, $\bar{\nu}_\mu$ CC quasi-elastic scattering $\bar{\nu}_\mu + p \to n + \mu^+$ can also contribute to this kind of background. 
It should be noted that this type of events was not observed by KamLAND \cite{KamLAND}. 
However, because of its larger target mass and proton exposure compared to KamLAND, it is possible for JUNO to observe these backgrounds. 
Since the energetic neutron usually breaks up the nucleus and produces many neutrons, a large number of neutron capture can be used to suppress this kind of background. 

\begin{figure}[h]
\centering
\includegraphics[width=8cm]{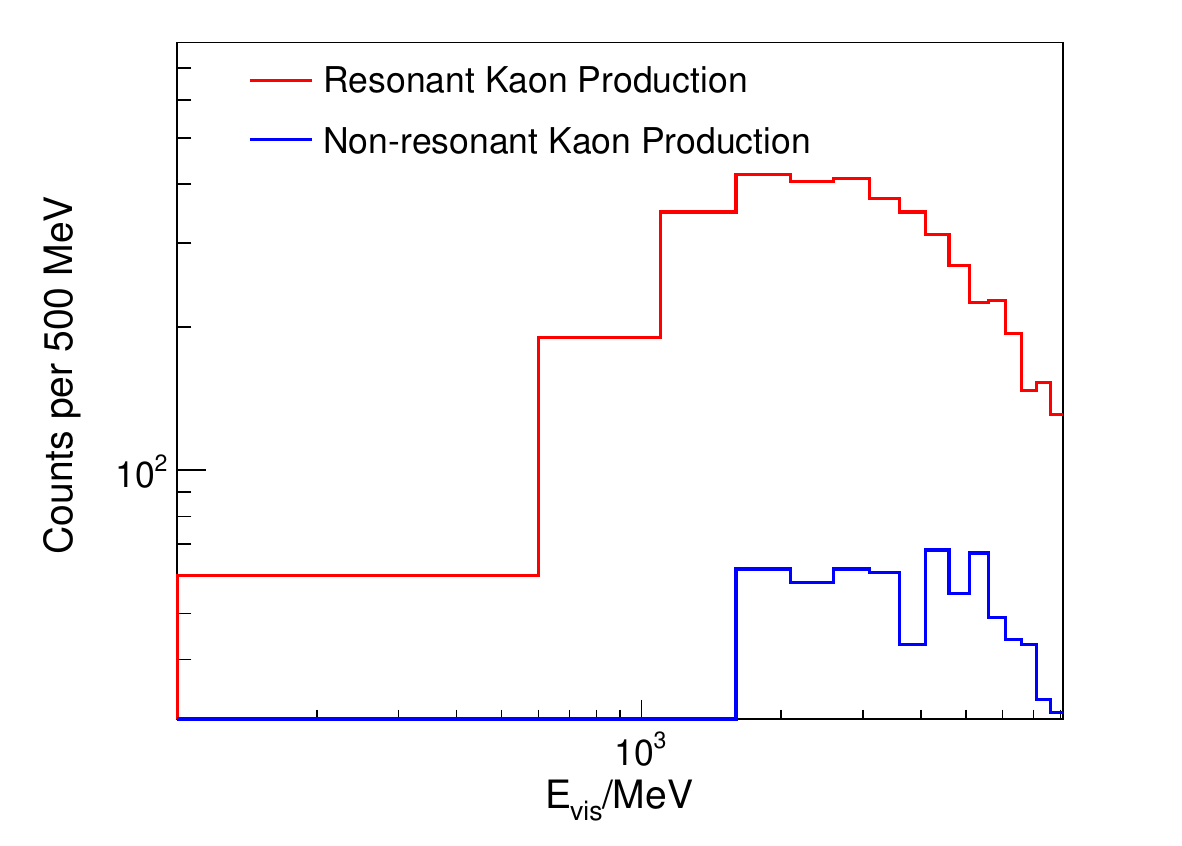}
\caption{\label{fig:KaonProd} Visible energy distribution of the kaon production from \AN backgrounds. According to the plot, the resonant kaon production has a negligible contribution and the non-resonant background can be eliminated, with an upper $E_\text{vis}$ cut at 600~MeV.}
\end{figure}

Another possible source of background is resonant and non-resonant kaon production (with or without $\Lambda$).
The visible energy distribution of the kaon is shown in Fig.~\ref{fig:KaonProd}. 
The Nuwro generator \cite{NuWro} is applied to help estimating the non-resonant kaon production, because this type of event is not included in GENIE due to the strangeness number conservation.
Based on the result of simulation, this kind of background has a negligible contribution in the relevant energy range (smaller than 600 MeV), which is similar to the LENA \cite{LENA} and KamLAND \cite{KamLAND} conclusions.

\section{Analysis} \label{Sec.4}

To quantify the performance of background discrimination, we design a series of selection criteria to evaluate the detection efficiency of \PD and the corresponding background rate based on the simulation data sample.
According to the physics mechanisms introduced in the last section, the key part of the selections is based on the triple coincidence signature in hit time spectrum. 
Many beneficial works to search for proton decay with a LS detector have been discussed by the LENA group and carried out by the KamLAND collaboration \cite{LENA,KamLAND}. 
However, the situation in JUNO is more challenging because of the much larger detector mass compared to KamLAND. 
Due to the relative masses, in ten years, the detected number of \AN would be about 20 times of that of the KamLAND experiment. 
Therefore, more stringent selection criteria have to be defined to suppress background to a level at least as low as that of KamLAND.
Besides the common cuts on energy, position and temporal features, additional criteria have to be explored. 
For the JUNO detector, a possible way to additionally distinguish the \PD is by using the delayed signals, including the Michel electron and neutron capture gammas.

\subsection{Basic Selections}

\begin{figure}
    \centering
        \label{EvisPD}
        \includegraphics[width=8cm]{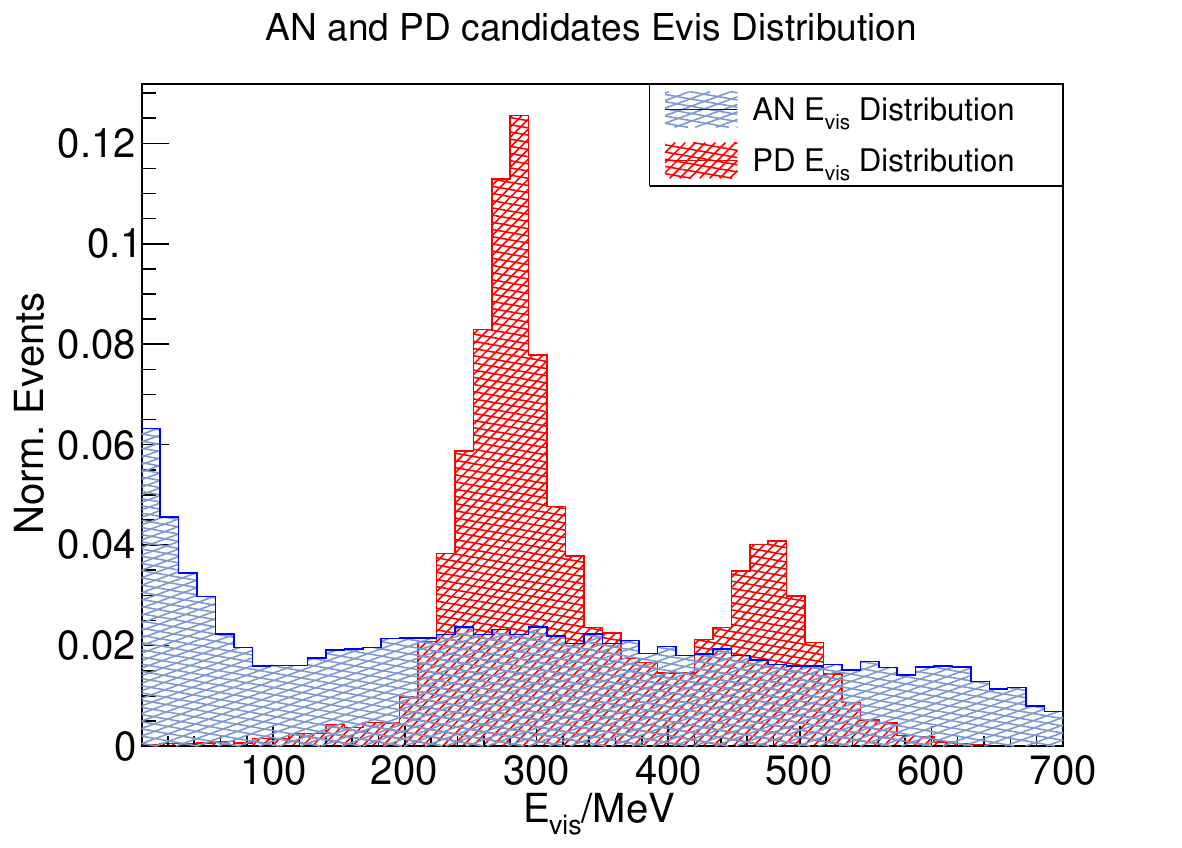}
    \caption{The visible energy distributions of \PD (PD) and \AN (AN) events. }
    \label{fig:EvisCom}
\end{figure}

The basic event selection uses only the most apparent features of the decay signature. 
The first variable regarded is the visible energy of the event.
The visible energy of \PD comes from the energy deposition of $K^+$ and its decay daughters.
The average energy deposition of $K^+$ is 105 MeV, while that of the decay daughters is 152 MeV and 354 MeV in the two dominant $K^+$ decay channels respectively.
Therefore, as illustrated in Fig.~\ref{fig:EvisCom}, the visible energy of \PD is mostly concentrated in the range of $200\, {\rm MeV} \leq E_\text{vis} \leq 600 \, {\rm MeV}$, comparable to that of the \AN backgrounds. 
Nearly half of the \AN events in the simulated event sample can be rejected with the $E_\text{vis} $ cut, while the \PD survival rate is more than $94.6\%$. 
The left and right peaks mainly correspond to the $K^+ \to \mu^+ \nu_\mu$ and $K^+ \to \pi^+ \pi^0$ decay channels, respectively. 

In the second step, if the CD is triggered, the VETO detector is required to be quiet in two consecutive trigger windows of 1000 ns which is before and after the prompt signals respectively.
In this way most muons can be removed, while the remaining muons usually get through the CD near its surface. 
The remaining muons usually have smaller visible energies and shorter track lengths. 
Thus, the track of the remaining muons should be closer to the boundary of the CD. 
Consequently, they can be further removed by a volume cut.
The volume within $R_V \leq 17.5$ m is defined as the fiducial volume of JUNO detector in \PD searches, thus the fiducial volume cut efficiency is 96.6\% and will be counted into the selection efficiency.

As shown in Table \ref{tab:eff}, after the basic cuts:
\begin{description}
 \item[(Cut-1)]  visible energy $200\, {\rm MeV} \leq E_\text{vis} \leq 600 \, {\rm MeV}$,
 \item[(Cut-2-1)] VETO system is not triggered in 1000 ns windows before and after the prompt signals,
 \item[(Cut-2-2)] volume cut is set as $R_V \leq 17.5$ m,
\end{description}
the survival rate of \PD in the simulated signal sample is $93.7\%$ while that of \AN events is 29.9\% from the total \AN events. Further selection methods to reduce the \AN background are required.

\subsection{Delayed Signals and Event Classification} \label{Sec:4.2}

Due to its good energy and time resolution, JUNO can measure the delayed signals of \PD and \AN events, including the Michel electron and neutron capture. 
About $95\%$ of \PD is followed by a Michel electron, while only 50\% of the background events exhibit a delayed signal after the basic selections. 
On the other hand, \PD on average has a smaller number of captured neutrons per event than the \AN events. 
Criteria can be set to further reduce the remaining background after the basic selection based on the differences between the characteristics of delayed signals. 

\begin{figure}[]
\centering
\subfigure[Distribution of $N_M$]{
    \includegraphics[width=8cm]{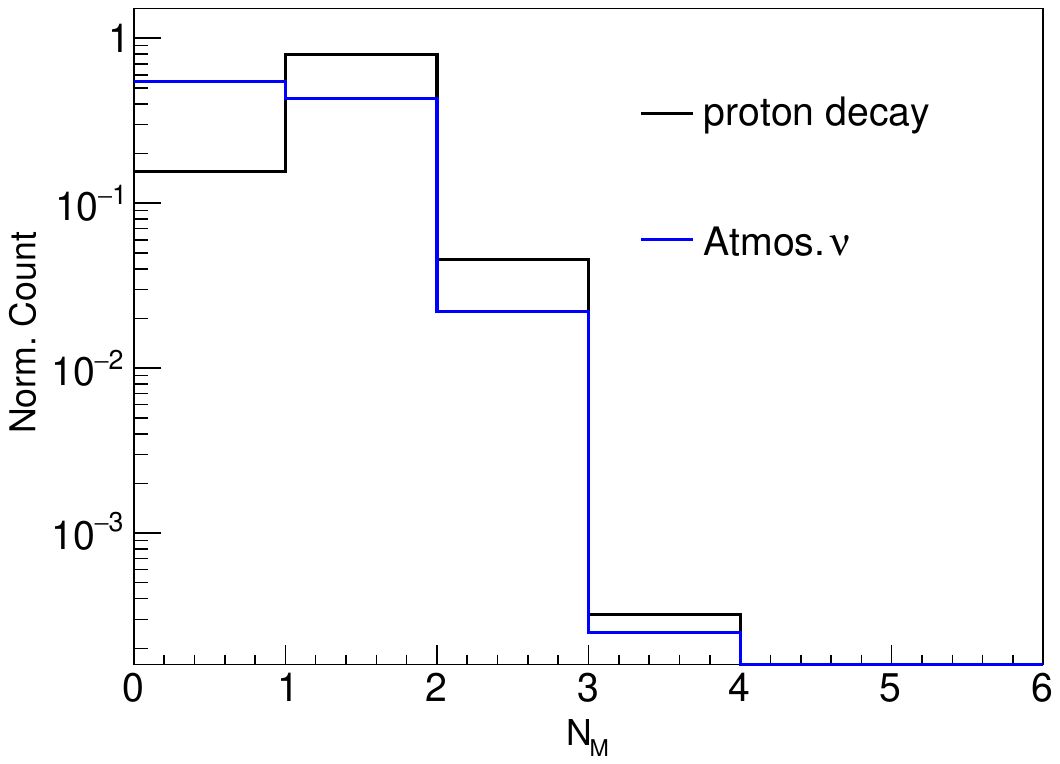}
}
\subfigure[Distribution of $\Delta L_{M}$]{
    \includegraphics[width=8cm]{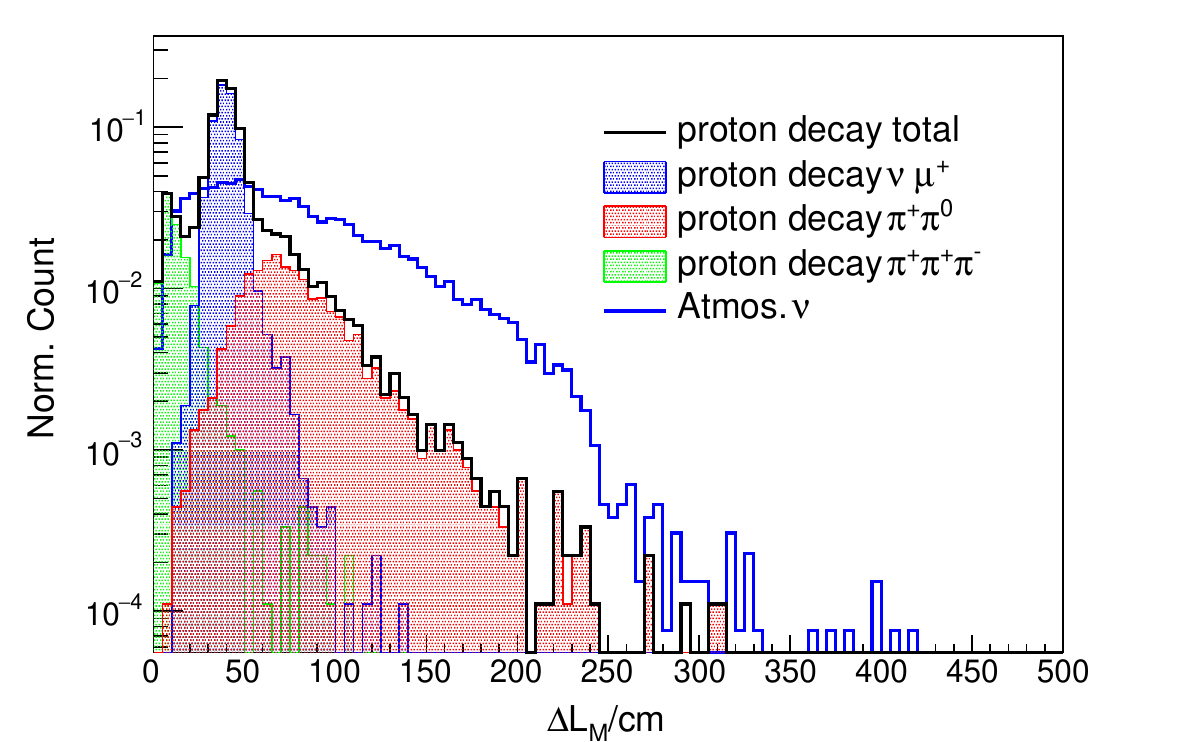}
}
\caption{\label{fig:Michel} The $N_M$ and $\Delta L_{M}$ distributions of identified Michel electrons for \PD and \AN events with the basic selection and the selection of the time and energy properties of Michel electrons. A unit area normalization is used.}
\end{figure}

The Michel electron is the product of the muon decay with kinetic energy up to 52.8 MeV and the muon lifetime is 2.2 $\mu$s.
For the Michel electron signals, we can know the visible energy $E_M$, the correlated time difference $\Delta T_{M}$ to the prompt signal and the correlated distance $\Delta L_{M}$ to the deposition center of the prompt signal from the MC simulation. 
Based on the physical properties of \PD and background events, it is assumed that JUNO can fully identify the Michel electron with $10\,{\rm MeV} < E_M < 54$~MeV and $150 \, {\rm ns} < \Delta T_{M} < 10000$~ns. In this case, the efficiency to distinguish Michel electrons is $89.2\%$. The lower limit of $E_M$ is set to avoid the influence of low energy background, like natural radioactivity. 
In Fig.~\ref{fig:Michel}, the number of events $N_{M}$ and $\Delta L_{M}$ distributions of identified Michel electrons for \PD and \AN events are shown.
About $5.58\%$ of the \PD events exhibit the number of Michel electrons $N_M =2$ which corresponds to the $K^+$ decay channel $K^+ \to \pi^+ \pi^+ \pi^-$.
For the $N_M =2$ case, $\Delta L_{M}$ is taken to be the average value of two correlated distances.
It is clear that proton decay has a smaller $\Delta L_{M}$ on average than the backgrounds. 
We can consequently use $\Delta L_{M}$ to reduce the \AN backgrounds by applying the criteria: 
\begin{description}
 \item[(Cut-3)] tagged Michel electron number $1 \leq N_M \leq 2$, 
 \item[(Cut-4)] correlated distance $\Delta L_{M} \leq 80$ cm,
\end{description}
in the remaining proton decay candidates after the basic selection. 
It can be found that $71.4 \%$ of \PD and $9.2 \%$ of \AN events survive in the simulated event samples.

\begin{figure}[]
\centering
\subfigure[Distribution of $N_n$]{
    \includegraphics[width=8cm]{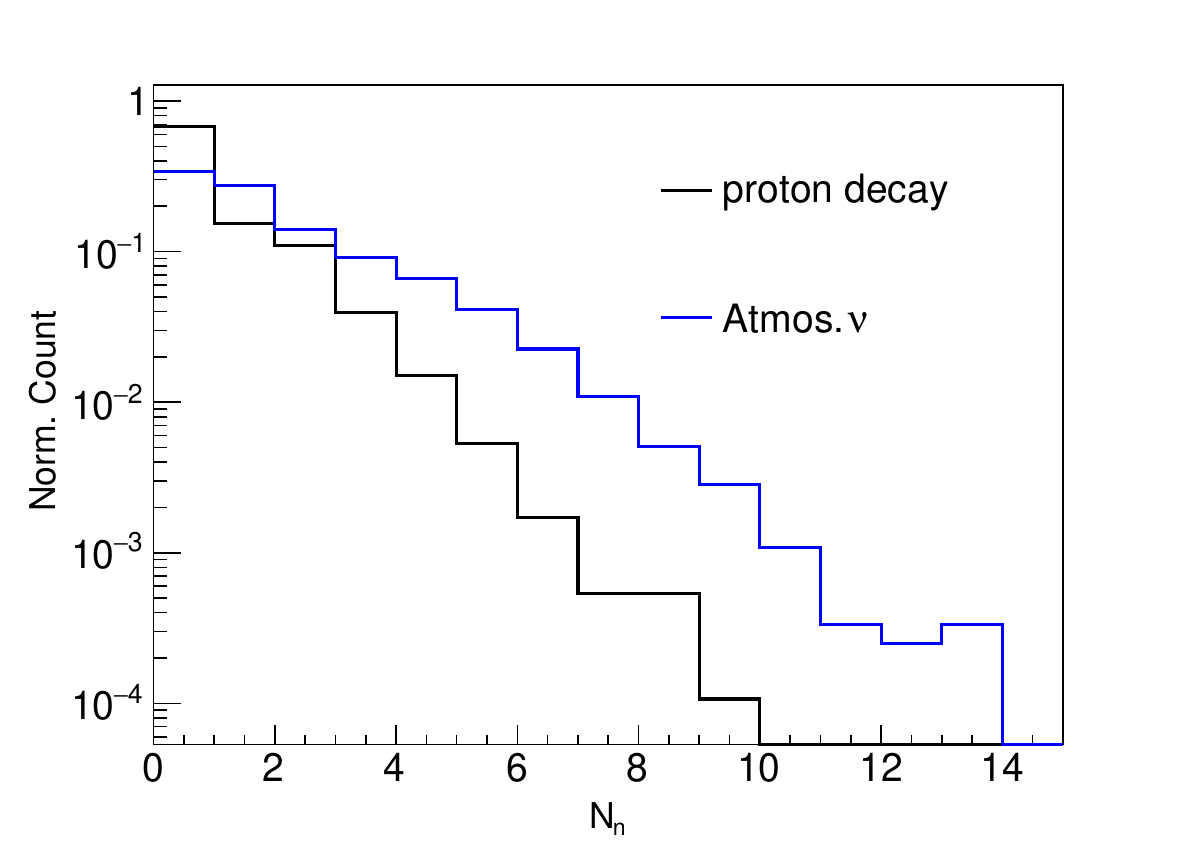}
}
\subfigure[Distribution of $\Delta L_{n}$]{
    \includegraphics[width=8cm]{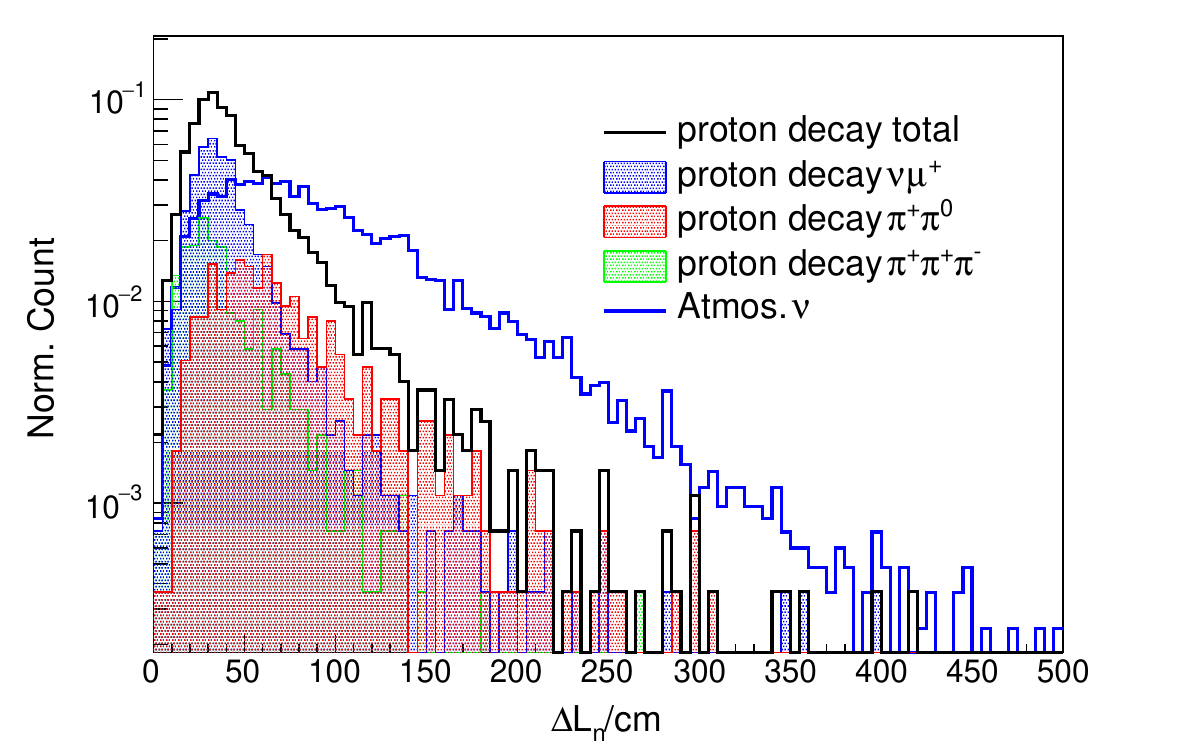}
}
\caption{\label{fig:Neutron} The $N_n$ and $\Delta L_{n}$ distributions of identified neutron capture for \PD and \AN events with the basic selection. A unit area normalization is used.}
\end{figure}

Similar to the Michel electron, the neutron capture is another potential selection criterion. 
Here we assume that the delayed neutron capture signal can be fully identified by requiring the visible energy $1.9 \,{\rm MeV} \leq E_n \leq 2.5$~MeV and the correlated time difference 1 $ \,{\rm \mu s}  \leq \Delta T_{n} \leq 2.5$~ms. 
In this way, $89.5\%$ of the neutrons produced by \AN events can be distinguished. 
In Fig. \ref{fig:Neutron}, the identified neutron distributions of \PD signals and backgrounds after the basic selections are shown. 
The proton decay events have a smaller $N_n$ on average than the \AN events. 
So we use the selection cut $N_n \leq 3$ to suppress the background. 
As shown in Fig. \ref{fig:Neutron}, the distance $\Delta L_{n}$, which is defined similarly to $\Delta L_{M}$, can also be a powerful tool to reduce the backgrounds.
Thus, a cut of $\Delta L_{n} \leq 70$ cm is required.
Note that these criteria about $N_n$ and $\Delta L_{n}$ can reduce an important class of background, namely events with a high energy neutron in the final state of the primary \AN interaction. 
Such a high energy neutron has a small probability not to lose its energy within 10~ns until it interacts with LS to give a second pulse. 
If the final states include $\mu^\pm$ or $\pi^+$, this background event will mimic the three fold coincidence of \PD. 
Since the high energy neutron usually produces more neutrons and larger $\Delta L_{n}$, we choose the following cuts:
\begin{description}
\item[(Cut-5)] tagged neutron number $ N_n \leq 3$ for $N_M=1$, 
\item[(Cut-6)] $\Delta L_{n} \leq 70$ cm if $N_M=1$ and $1 \leq N_n \leq 3$,
\end{description}
to suppress this kind of background.

Based on the above discussions about the delayed signals, we naturally classify the MC events into the following three samples:
\begin{description}
\item[Sample 1] $N_M=1,\Delta L_{M} \leq \SI{80}{cm}, N_n = 0$;
\item[Sample 2] $N_M=1,\Delta L_{M} \leq \SI{80}{cm}, 1 \leq N_n \leq 3, \Delta L_{n} \leq \SI{70}{cm}$;
\item[Sample 3] $N_M=2, \Delta L_{M} \leq \SI{80}{cm}$.
\end{description}
The survival rate of \PD and the atmospheric $\nu$ events in the simulation can be found in Table \ref{tab:eff}. About 6.8\% of the total atmospheric $\nu$ events would survive, requiring further selection methods to reduce the background.

\subsection{Multi-Pulse Fitting}
\label{multipulsefitting}

\begin{figure*}[]
\centering
\subfigure[Hit time spectrum of a proton decay event]{
    \includegraphics[width=8cm,trim=0 0 0 30,clip]{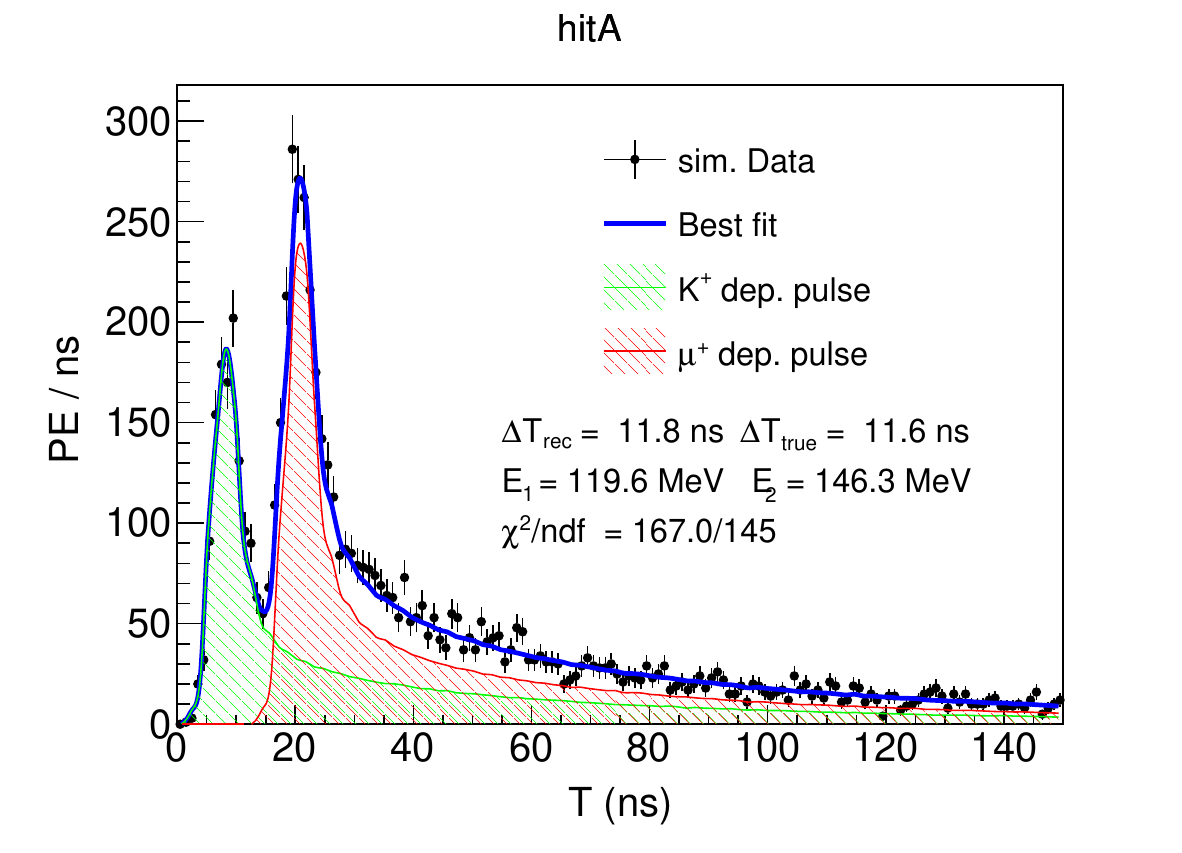}
}
\subfigure[Hit time spectrum of an \AN event]{
    \includegraphics[width=8cm,trim=0 0 0 30,clip]{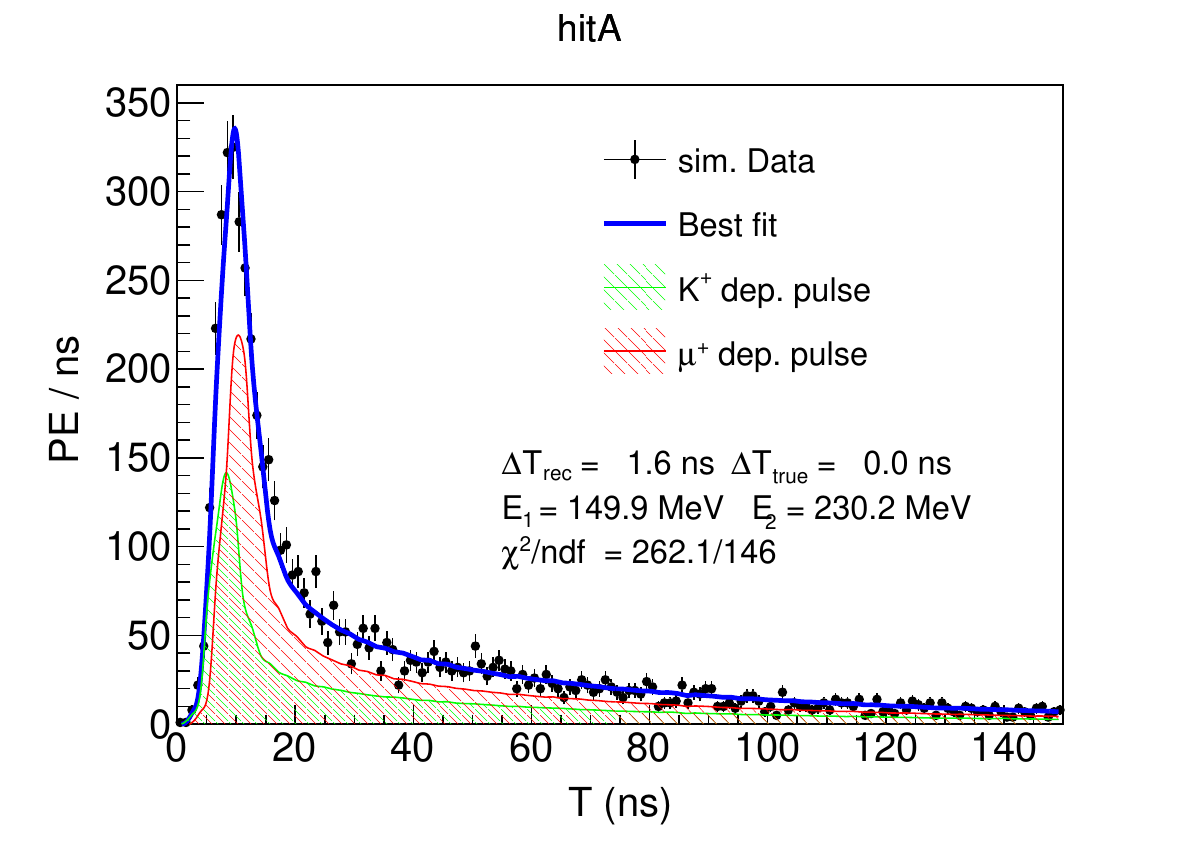}
}
\caption{\label{fig:fitdemo} Illustration of multi-pulse fitting to hit time spectra of a proton decay event (left) and an \AN event (right). The x axis is the hit time after TOF correction. The black dots are the observed spectrum from simulation. The blue line is the fitting result. The green and red filled histograms are the fitted result of the two components in the hit time spectrum which are contributed by the $K^+$ and the $K^+$ decay daughters.}
\end{figure*}

As introduced in Sec. \ref{sec:PD}, a \PD event usually has a triple coincidence signature on its hit time spectrum. 
The first two pulses of the triple coincidence overlap with each other concerning the decay time of \KPLUS, which is a distinctive feature of \PD comparing to the \AN backgrounds. 
It means that the \PD can be distinguished from the backgrounds according to the characteristics of the overlapping double pulses. 
Therefore, the hit time spectrum is studied further by multi-pulse fitting method \cite{KamLAND}, in order to reconstruct the time difference and energy of the $K^+$ and its decay daughters.

For each event, its hit time spectrum can be fitted with double-pulse $\phi_{D}(t)$ and single-pulse $\phi_{S}(t)$ templates of hit time $t$,
\begin{eqnarray}
        \phi_{D}(t;\epsilon_{K}, \epsilon_{i}, a, \Delta T) & = &\epsilon_{K} \phi_{K}(t) + \epsilon_{i} \phi_{i}[a(t-\Delta T)],  \label{phi_D} \\
        \phi_{S}(t;\epsilon_{S}) & = & \epsilon_{S} \phi_\text{AN}(at),  \label{phi_S} 
        \end{eqnarray}
where $\phi_{K}(t)$ is the TOF-corrected template of \KPLUS, $\phi_{i}(t)$ is that of a decay daughter of \KPLUS. $i = \mu$ and $\pi$ refer to the two dominant decay channels $K^+ \to \mu^+ \nu_\mu$ for $E_\text{vis}\leq \SI{400}{MeV}$ and $K^+ \to \pi^+ \pi^0$ otherwise. These templates are produced by the MC simulations in which the particles are processed by SNiPER with their corresponding kinetic energies. $\phi_\text{AN}(t)$ is the template of the backgrounds, generated as the average spectrum of all the \AN events with energy deposition from \SIrange{200}{600}{MeV}.  Due to the influence of reflection, the hit time spectrum is widened when the energy deposition center is close to the boundary. In order to deal with this effect, the templates are separately produced in inner volume ($<$ 15 m) and outer volume ($>$ 15 m), and applied to the fitting of events in the corresponding volumes respectively.

In Eqs. (\ref{phi_D}) and (\ref{phi_S}), $\Delta T$ is the correlated time difference of the delayed component, $a$ is a scaling factor to account for shape deformation of the second pulse caused by the electromagnetic showers, and $\epsilon _{K}$, $\epsilon _{i}$, $\epsilon_{S}$ are the corresponding energy factors.  They are free parameters in the fitting. For illustration, we use Eq.~(\ref{phi_D}) to fit two typical events as shown in Fig.~\ref{fig:fitdemo}.

After fitting the hit time spectra with the templates of Eqs. (\ref{phi_D}) and (\ref{phi_S}), we calculate the $\chi^2$ of the double and single pulse fittings using
\begin{eqnarray}
        \chi_D^2&=& \sum \frac{[\phi(t)-\phi_D(t)]^2}{\sigma^2 [\phi(t)]}\label{eq:chi2-gaussian}, \\
        \chi_S^2&=& \sum \frac{[\phi(t)-\phi_S(t)]^2}{\sigma^2 [\phi(t)]},
\end{eqnarray}
where \(\sigma^2 [\phi(t)]\) is the sample variance of the observed spectrum \(\phi(t)\)  at the \(t\)-th bin. The $\chi ^2$ ratio $R_{\chi} \equiv \chi_S^2/\chi_D^2$ is taken as the further selection criterion. 
From the double-pulse fitting by Eq. (\ref{phi_D}), the energies $E_{1}$ and $E_{2}$ of the overlapping double pulses from depositions of the postulated $K^+$ and its decay daughters are calculated from \(\epsilon_K\), \(\epsilon_i\) and \(a\) introduced in Eq. (\ref{phi_D}),
\begin{eqnarray}
        E_{1}&=&\frac{\epsilon_K T_{K}}{\epsilon_K T_{K} + \epsilon_i T_{i}/a} E_\text{fit}    \label{E1} \\
        E_{2}&=&\frac{\epsilon_i T_{i}/a}{\epsilon_K T_{K} + \epsilon_i T_{i}/a} E_\text{fit},  \label{E2}
\end{eqnarray}
where $T_K = 105$ MeV is the initial kinetic energy of $K^+$ from the free proton decay. $T_\mu = 152$ MeV and $T_\pi = 354$~MeV are the initial kinetic energies of muon and pion from the $K^{+}$ decay at rest. The fitted total energy is defined as $E_\text{fit} = E_\text{vis} - \sum E_M - \sum E_n$ which is the visible energy subtracting the energies of Michel electrons and neutron captures.

\begin{figure}[h]
\includegraphics[width=8cm]{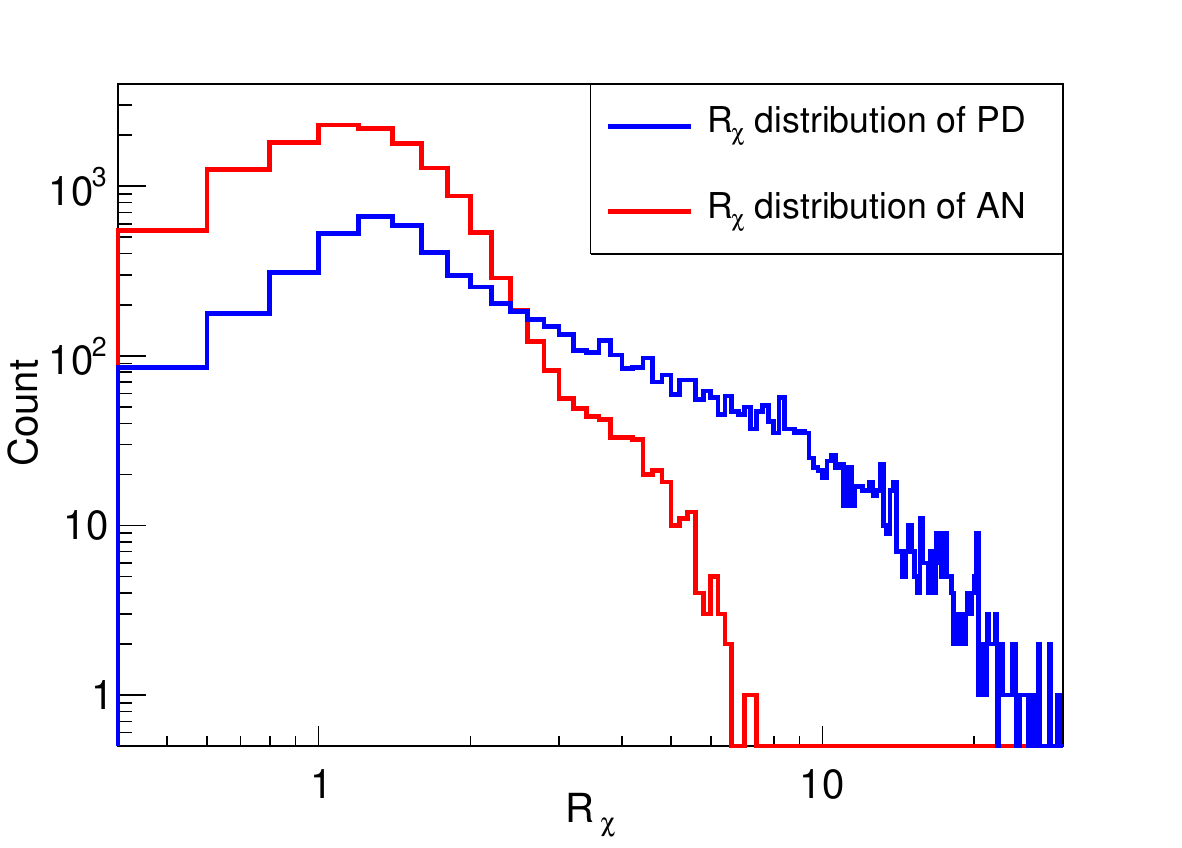}
\caption{\label{fig:chi2Ratio} Distributions of the $\chi ^2$ ratio $R_{\chi} \equiv \chi_S^2/\chi_D^2$ from the \PD (PD) and \AN (AN) events after the basic selection and the delayed signal selection.}
\end{figure}

The way to select \PD from the \AN backgrounds according to the parameters acquired above will be introduced as follows. 
In Fig. \ref{fig:chi2Ratio}, we plot $R_{\chi}$ distributions for the proton decay and the \AN events after applying the selections from Cut-1 to Cut-6. 
It can be found that $R_{\chi}$ is a tool to reject the background. 
Actually, the $R_{\chi}$ can be regarded as an indicator that the fitted event tends to be a double pulse overlapping event or a single pulse event. 
The larger the $R_{\chi}$ is, the stronger it tends to be an event with two pulses overlapping in hit time spectrum.
A cut of $R_{\chi} >1$ can be applied to roughly do the selection. 
If $R_{\chi} >1$, this fitted event could be preliminarily identified as a proton decay candidate. Otherwise, it would be rejected as a background candidate.
However, a general cut of the $R_{\chi}$ is not justified to the three samples defined at the end of Sec. \ref{Sec:4.2}. 
Compared to sample 1 which is composed of the common \PD and \AN events, sample 2 is additionally composed of the background events with energetic neutrons introduced in Sec.~\ref{sec:AN}. 
The second pulse caused by an energetic neutron makes these \AN events have a fake double pulse overlapping shape in the hit time spectrum. 
A stricter requirement to the $R_{\chi}$ is consequently necessary to reduce the background. 
The $K^+$ produced in the \PD events in sample 3 actually decays via $K^{+}\to\pi^{+}\pi^{+}\pi^{-}$ due to the cut on the number of Michel electrons $N_{M}=2$. 
As a result, the \PD should be easier to be distinguished from the backgrounds with $N_{M}=2$. 
Therefore it is reasonable to set a less stringent cut on $R_{\chi}$ in order to keep a high detection efficiency. Consequently, the $R_{\chi}$ will be set for the three samples separately.
In order to sufficiently reject \AN backgrounds, we require
\begin{description}
\item[(Cut-7-1)]  $R_\chi > 1.1$ for Sample 1,
\item[(Cut-7-2)]  $R_\chi > 2.0$ for Sample 2,
\item[(Cut-7-3)]  $R_\chi > 1.0$ for Sample 3.
\end{description}

\begin{figure}[h]
\centering
\subfigure[Distribution of fitted $\Delta T$]{
    \includegraphics[width=8cm]{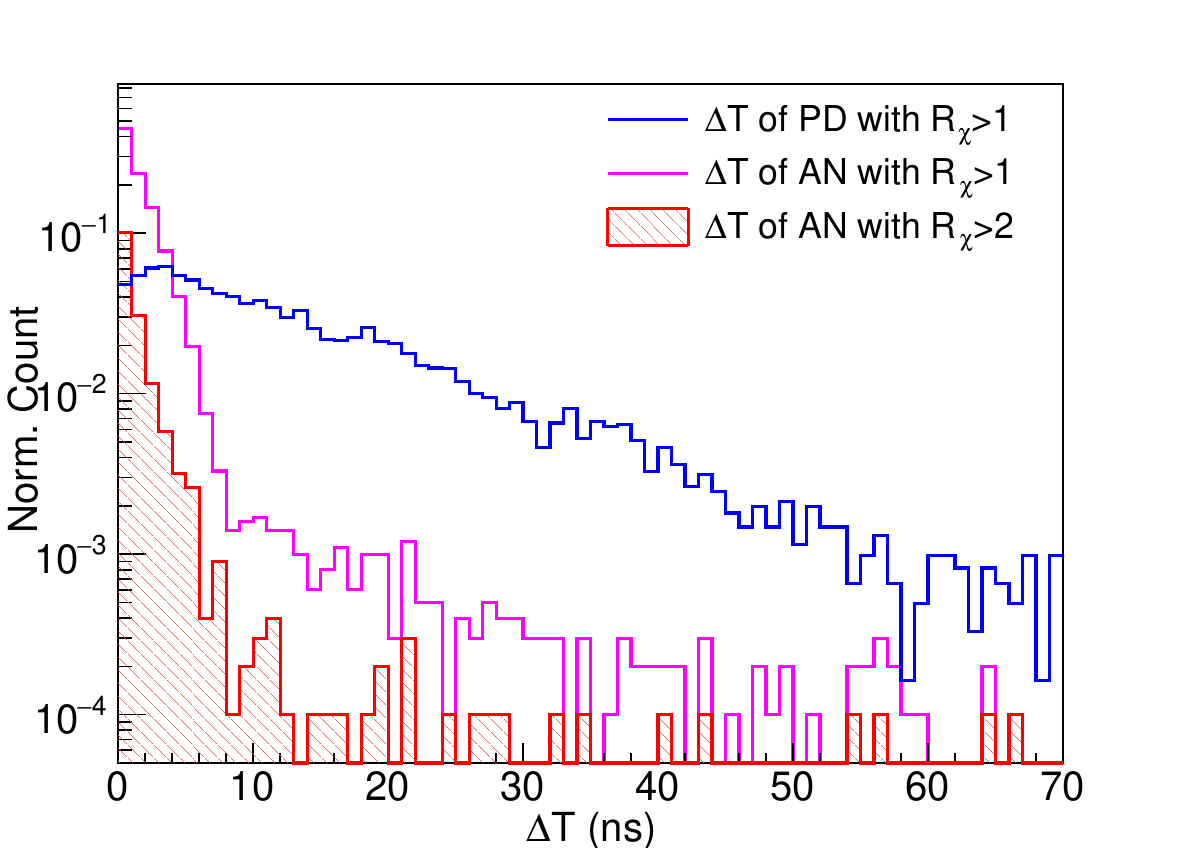}
}
\subfigure[Fitting efficiencies of \PD]{
    \includegraphics[width=8.3cm]{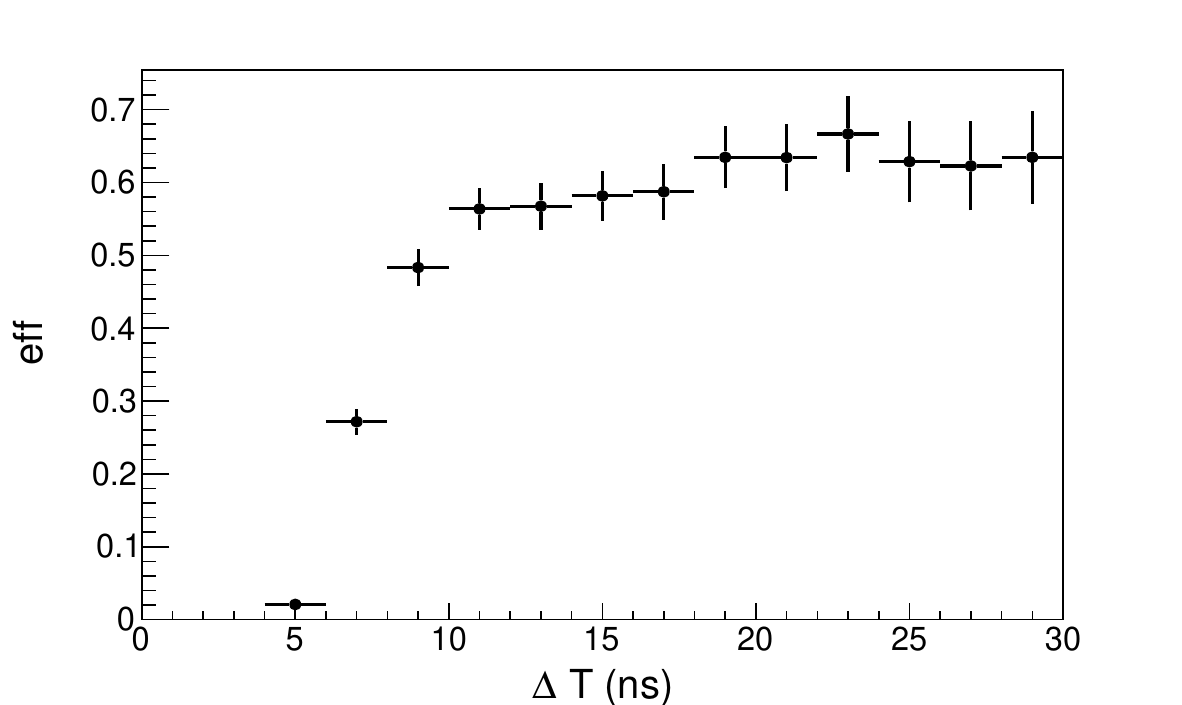}
}
\caption{\label{fig:FitTime} $\Delta T$ distribution and Fitting efficiencies. (a) Distribution of fitted $\Delta T$ (equation (\ref{phi_D})) of \PD (PD, in blue)  and \AN (AN, red filled and pink) events with different $R_{\chi}$ cuts after the basic selection and delayed signal selection. (b) Fitting efficiencies for \PD with different true $\Delta T$ ($K^+$ decay time). The efficiencies are low when $K^+$ decays within several ns because both pulse components are too close.}
\end{figure}

The distributions of fitted $\Delta T$ are shown in Fig.~\ref{fig:FitTime}(a), where a rough cut of $R_{\chi}>1$ is applied to \PD and the backgrounds. 
From the figure, it can be found that $\Delta T$ for the remaining backgrounds which are mis-identified as \PD candidates are mostly distributed at small $\Delta T$, because the \AN events are usually a single pulse. 
Meanwhile, when the $K^+$ decays in few nanoseconds, the fitting has low efficiency because both components are too close to be distinguished from each other (as Fig.~\ref{fig:FitTime}(b) shown). 
Consequently, $\Delta T$ is required as:
\begin{description}
 \item[(Cut-8)] correlated time difference should be $\Delta T \geq 7$ ns,
\end{description}

\begin{figure*}[]
\centering
\subfigure[\PD]{
    \includegraphics[height=10cm]{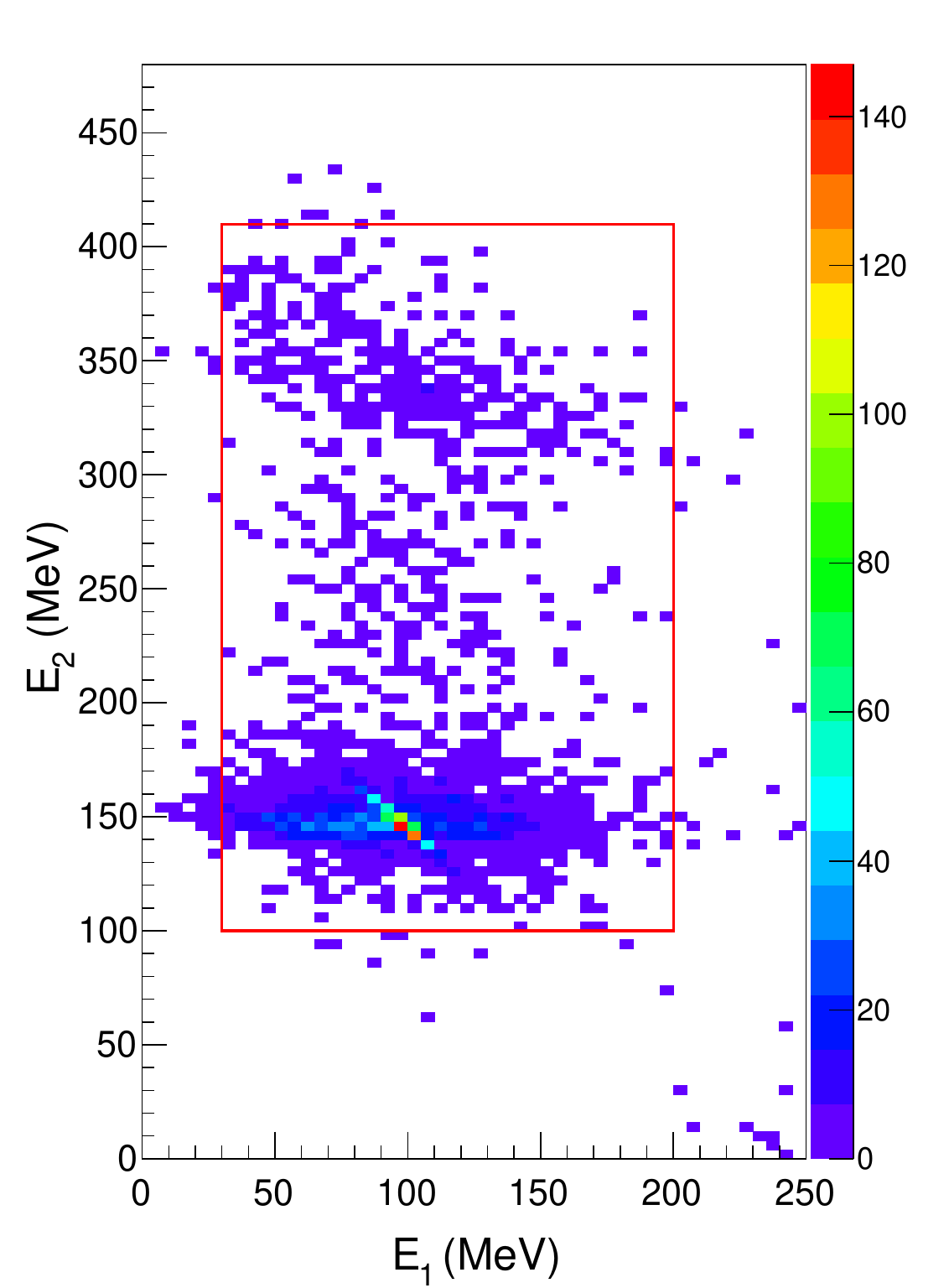}
}
\subfigure[\AN]{
    \includegraphics[height=10cm]{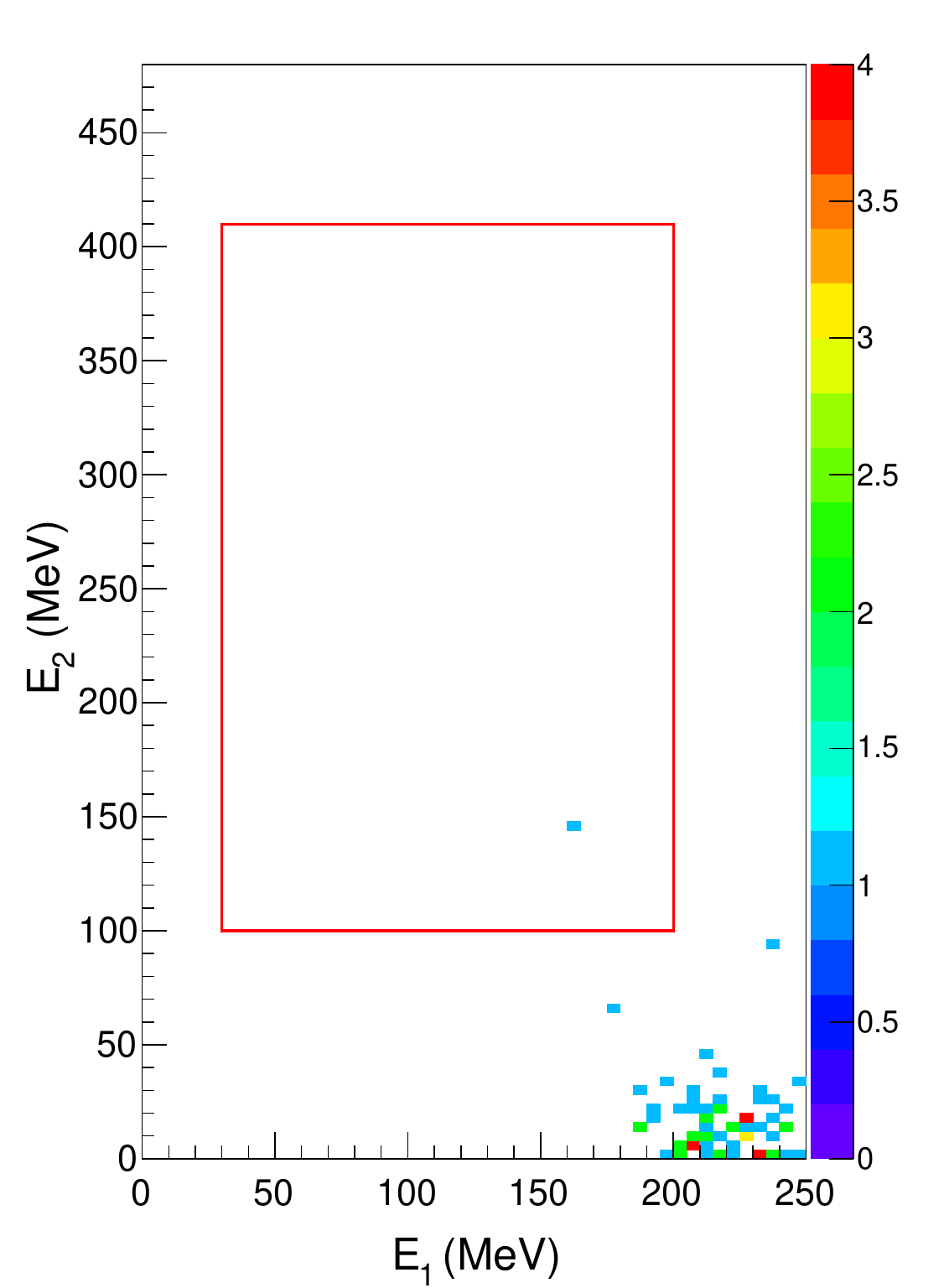}
}
\caption{\label{fig:E1E2} Correlated $E_1$ and $E_2$ distributions (in colored scale) for the \PD (a) and \AN (b) events with the basic selection, delayed signal selection, the $R_{\chi}$ cut and the $\Delta T$ cut. The events out of the red boxes would be rejected as the background. More details can be found in the text.}
\end{figure*}

Concerning the kinematics of the $K^+$ and its decay daughters, the sub-energy $E_1$ should be distributed from 0 to more than 200 MeV with an average of 105 MeV, while $E_2$ should be fixed around 152 MeV or 354 MeV depending on the decay mode.
As shown in Fig.~\ref{fig:E1E2}, we plot the correlated sub-energy deposition distributions of \PD and background events.
Two obvious groups in the left panel can be observed, corresponding to the two dominant decay channels of $K^+$.
Only a small group of \AN events is left in the bottom right corner of the right panel of Fig.~\ref{fig:E1E2}, which comes from the mis-identification of a tiny second peak. 
It is clear that a box selection on $E_1$ and $E_2$ can efficiently reject the \AN backgrounds. Therefore the selections, 
\begin{description}
\item[(Cut-9-1)]  $30 \, {\rm MeV} \leq E_1 \leq 200$ MeV 
\item[(Cut-9-2)]  $100 \, {\rm MeV} \leq E_2 \leq 410$ MeV, 
\end{description}
are required. The lower boundary of $E_1$ is set to avoid the influence of the coincidence with the low energy events like reactor antineutrinos or radioactive backgrounds. 

\begin{table*}[]
\centering
\caption{\label{tab:eff} Detection efficiencies of \PD and the number of \AN background after each selection criterion. The total amount of atmospheric $\nu$ background simulated is 160 k, which corresponds to an exposure of 890 kton-years.}
\begin{tabular}{c|c|c|c|c|c|c|c}
\Xhline{1.2pt}
\multicolumn{2}{c|}{\multirow{2}{*}{Criteria}}   & \multicolumn{3}{c|}{Survival rate of \PD (\%)} & \multicolumn{3}{c}{ \ \ \ \ \ \  Survival count (fraction) of \AN \ \ \ \ \ \  } \\  \cline{3-8}
\multicolumn{2}{c|}{}                            & Sample 1  & Sample 2  & Sample 3  & Sample 1  & Sample 2   & Sample 3 \\  \hline
\multirow{2}{*}{\tabincell{c}{basic selection}}  
      & $E_\text{vis}$ & \multicolumn{3}{c|}{94.6}   & \multicolumn{3}{c}{51299  (32.1\%)}  \\\cline{2-8}
      & $R_V$         & \multicolumn{3}{c|}{93.7}   & \multicolumn{3}{c}{47849  (29.9\%)}  \\ \hline
\multirow{4}{*}{\tabincell{c}{Delayed \\signal \\selection}} 
      & $N_M$      & \multicolumn{2}{c|}{74.4}   & 4.4   & \multicolumn{2}{c|}{20739  (13.0\%)}   & 1143  (0.7\%)  \\  \cline{2-8}
      & $\Delta L_M$     & \multicolumn{2}{c|}{67.0}   & 4.4   & \multicolumn{2}{c|}{13796  (8.6\%)}   & 994  (0.6\%)   \\  \cline{2-8}
      & $N_n$      & 48.4                       & 17.9  & --     & 5403  (3.4\%)   & 6857  (4.3\%)      & --    \\  \cline{2-8}
      & $\Delta L_n$     & --                          & 16.6  & --     & --     & 4472  (2.8\%)      & --               \\  \hline
\multirow{3}{*}{\tabincell{c}{Time \\character \\selection}}
      & $R_{\chi}$ & 45.9                      & 9.0   & 3.8   & 4326  (2.7\%)   & 581  (0.4\%)       & 716  (0.4\%)        \\  \cline{2-8}
      & $\Delta T$ & 28.3                      & 7.7   & 2.4   & 121  (0.07\%)    & 18  (0.01\%)        & 30  (0.02\%)             \\ \cline{2-8} 
      & $E_{1}, E_{2}$   & 27.4                 & 7.3   & 2.2   & 1  (0.0006\%)      & 0         & 0              \\  \hline
\multicolumn{2}{c}{Total}                 & \multicolumn{3}{c}{36.9}                               & \multicolumn{3}{c}{1}  \\ \Xhline{1.2pt}                                  
\end{tabular}
\end{table*}

The detection efficiencies under each selection criterion are listed in Table \ref{tab:eff}, where the numbers of the remaining backgrounds are also shown, from which the elimination power of each criterion can be found. 
After applying these criteria, the total efficiency for \PD is estimated to be $36.9\%$, while only one event in sample 1 remains from the simulated 160 k \AN events (corresponding to an exposure of 890 kton-years or exposure time of 44.5 years on JUNO site).
Since the volume cut in the basic selections provides a selection efficiency of 96.6\% to the total efficiency, it will not be counted in the exposure mass calculation.
The three samples contribute to $27.4\%$, $7.3\%$ and $2.2\%$ of the detection efficiencies, respectively.   
Considering the statistical error and the weighting value which accounts for the oscillation probability, the background level corresponds to 0.2 events which has been scaled to 10 years data taking of JUNO.

\section{Sensitivities and Uncertainties } \label{Sec.5}

The detection efficiency uncertainties of \PD are estimated in Table \ref{tab:effuncertainty}. 
The statistical uncertainty is estimated to be 1.6\% in the MC simulation. 
So far, we are using the ideal setting for the position reconstruction (30~cm of the energy deposition center position uncertainty without bias).
Considering the performance of the vertex reconstruction algorithm, it is assumed that the residual bias of the position reconstruction of \PD is 10~cm. 
In this case, the efficiency uncertainty caused by the volume cut of 17.5 m will be 1.7\%.

\begin{table}[]
\centering
\caption{\label{tab:effuncertainty} The detection efficiency uncertainties for \PD.}
\begin{tabular}{cc}
\Xhline{1.2pt}
Source                  & Uncertainty \\ \hline
Statistic               & 1.6\%      \\ \hline
Position reconstruction & 1.7\%       \\ \hline
Nuclear model           & 6.8\%      \\ \hline
Energy deposition model & 11.1\%      \\ \hline
Total                   & 13.2\%    \\ \Xhline{1.2pt}
\end{tabular}
\end{table}

Another important systematic uncertainty of detection efficiency comes from the inaccuracy of the nuclear model which will influence the ratio of accompanying particles of \PD. 
To estimate this uncertainty, another \PD sample base is simulated with the FSI and de-excitation processes of the residual nucleus disabled. 
After applying all the criteria, a difference in the detection efficiency is found to be 6.8\%, which is the estimation of the uncertainty from the nuclear model.

The dominant uncertainty comes from the energy deposition model. 
Due to the lack of study on Sub-GeV particles' behavior, especially the quenching effect of hundreds of MeV $K^+$ in LAB based LS, the deposition simulation in the LS detector might be inaccurate. 
Therefore, the simulated waveform of the hit time spectrum might be different from the real one. 
According to the study of KamLAND \cite{KamLAND}, this kind of uncertainty is estimated as $11.1\%$. 
We conservatively use this value considering the similar detection method. 
Therefore, the uncertainty of the proton lifetime is estimated as $13.2\%$ considering all the sources introduced above.

The uncertainties of the background level in ten years is composed of two parts. 
One is the systematic uncertainty that is contributed by the uncertainty of the atmospheric neutrino flux (20\%) and the atmospheric neutrino interaction cross-section (10\%) \cite{JUNO}. 
Another uncertainty comes from the number $N_n$ of neutron captures, which can be affected by the secondary interactions of hadronic daughter particles of atmospheric neutrino events in the LS. 
This is estimated as 10\% assuming the same uncertainty as Super-K \cite{Super-Kamiokande:2016exg}. 
The statistic uncertainty is estimated following the $1/\sqrt{N}$ rule. 
Considering that only one event survives in the selection, it is calculated as $\pm 0.2$ in ten years. 
With 160 k events in the current MC simulation, it is hard to improve since it will consume vast computing resources. We hope to update this value with a larger MC simulation data volume when it permits. Consequently, the background is estimated as $0.2\pm 0.05({\rm syst})\pm 0.2({\rm stat})$.

\begin{figure}[]
    \includegraphics[height=6.5cm]{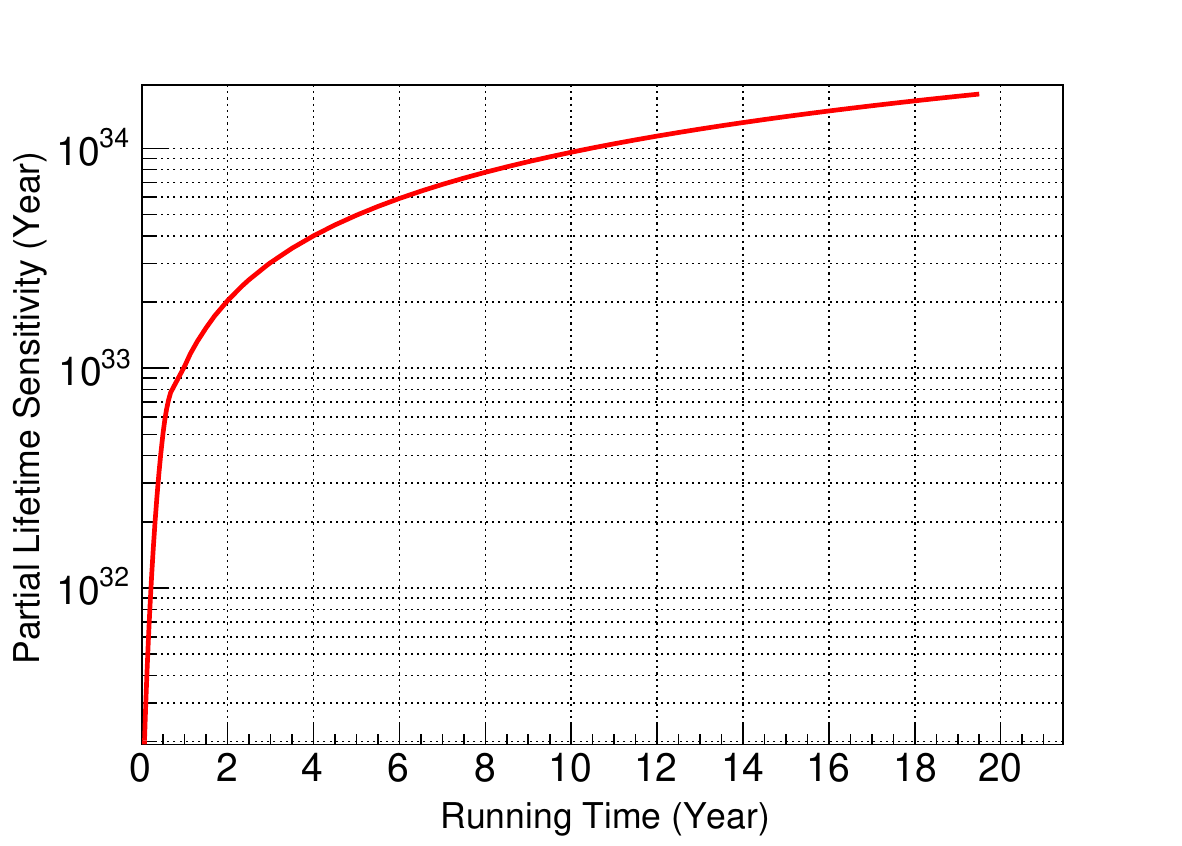}
\caption{\label{fig:sensitivity} JUNO sensitivity for \PD as a function of running time.}
\end{figure}

The sensitivity on \PD is expressed as    
\begin{eqnarray}
\tau / B(p \to \bar{\nu} K^+) = \frac{N_p T \epsilon}{n_{90}},
\end{eqnarray}
where $N_p = 6.75 \times 10^{33}$ is the total number of protons (including $1.45\times 10^{33}$ free protons and $5.3\times 10^{33}$ bound protons) in the JUNO central detector, $T$ is the running time which is assumed to be 10 years to achieve exposure mass of 200 kton-years, $ \epsilon =36.9\%$ is the total signal efficiency. $n_{90}$ is the upper limit of $90\%$ confidence level of the detected signals. 
It depends on the number of observed events and background level. 
According to the Feldman-Cousins method \cite{Feldman:1997qc}, $n_{90}$ is estimated as 2.61 given an expected background of 0.2 in 10 years. 
Thus, the JUNO sensitivity on \PD at 90\% C.L. with 200 kton-years would be
\begin{equation}
\tau / B(p \to \bar{\nu} K^+) > 9.6 \times 10^{33} \, {\rm years}. \label{eq:sensitivity} 
\end{equation}
Comparing to the representative liquid scintillator detector, the detection efficiency on \PD of JUNO is relatively lower than LENA \cite{LENA}. 
This should be reasonable considering that the study is based on an overall detector simulation of JUNO. 
Based on the background level 0.02 events per year, JUNO sensitivity as a function of running time is plotted as shown in Fig.~\ref{fig:sensitivity}. 
After 6 years running (120 kton-years), JUNO will overtake the current best limit from the Super-K experiment.

Moreover, the proton lifetime measured by JUNO will reach $10^{34}$ years for the first time after data taking of 10.5 years. 
In the case of no event observation after ten years, the $90\%$ C.L. limit to the proton lifetime would reach $1.1 \times 10^{34}$ years. 
In the case of one event observation ($16.4\%$ probability), the corresponding limit would be $6.0 \times 10^{33}$ years.

\section{Conclusion} \label{Sec.6}

A Simulation study to estimate the performance of the JUNO detector on searching for proton decay via \PD has been presented. 
It is found that the expected detection efficiency of \PD is $36.9\%\pm4.9\%$, while the background is estimated to be $0.2\pm 0.05({\rm syst})\pm 0.2({\rm stat})$ in ten years exposure. 
Assuming no proton decay events observed, the sensitivity of JUNO on \PD is estimated to be $9.6 \times 10^{33}$ years at 90\% C.L. based on the total exposure of 200 kton-years (or a live fiducial exposure of 193 kton-years). 
This is higher than the current best limit $5.9\times10^{33}$ years from the excellent effort of Super-K experiment with a live fiducial exposure of 260 kton-years \cite{Abe:2014mwa}. 

It shows that a liquid-scintillator detector like JUNO will be competitive when compared to the planned Hyper-Kamiokande \cite{Hyper-K} and DUNE \cite{DUNE} experiments.
Using different target nuclei $^{12} {\rm C}$ from the liquid scintillator and the newly developed analysis method considering the delayed signals (the Michel electrons and neutron captures), JUNO will provide a complementary search to test the GUTs from the view of \PD.
Besides the \PD mode, JUNO will have some sensitivity to the other nucleon decay modes listed in Ref. \cite{PDG}, particularly to the decay modes that also have the three fold coincidence feature in time, such as $n\to \mu^- K^+$, $p\to e^+K^{*}(892)^{0}$, $n\to \nu K^{*}(892)^0$ and $p\to \nu K^{*}(892)^{+}$.
They will be analyzed in the future.

\section*{Acknowledgement}

We are grateful for the ongoing cooperation from the China General Nuclear Power Group.
This work was supported by
the Chinese Academy of Sciences,
the National Key R\&D Program of China,
the CAS Center for Excellence in Particle Physics,
Wuyi University,
and the Tsung-Dao Lee Institute of Shanghai Jiao Tong University in China,
the Institut National de Physique Nucl\'eaire et de Physique de Particules (IN2P3) in France,
the Istituto Nazionale di Fisica Nucleare (INFN) in Italy,
the Italian-Chinese collaborative research program MAECI-NSFC,
the Fond de la Recherche Scientifique (F.R.S-FNRS) and FWO under the ``Excellence of Science - EOS'' in Belgium,
the Conselho Nacional de Desenvolvimento Cient\'ifico e Tecnol\`ogico in Brazil,
the Agencia Nacional de Investigacion y Desarrollo in Chile,
the Charles University Research Centre and the Ministry of Education, Youth, and Sports in Czech Republic,
the Deutsche Forschungsgemeinschaft (DFG), the Helmholtz Association, and the Cluster of Excellence PRISMA+ in Germany,
the Joint Institute of Nuclear Research (JINR) and Lomonosov Moscow State University in Russia,
the joint Russian Science Foundation (RSF) and National Natural Science Foundation of China (NSFC) research program,
the MOST and MOE in Taiwan,
the Chulalongkorn University and Suranaree University of Technology in Thailand,
and the University of California at Irvine in USA.

\end{document}